%% file: main.tex
\documentclass[sigconf]{acmart}

\renewcommand\footnotetextcopyrightpermission[1]{} % removes footnote with conference information in first column
\setcopyright{none}
\usepackage[normalem]{ulem}

\usepackage{amsmath,amsfonts}
\usepackage{algpseudocode}
\usepackage{algorithm}
\usepackage{textcomp}
\usepackage{xcolor}
\usepackage{xspace}
\usepackage{balance}
\usepackage{enumitem}
\usepackage{multirow}
\usepackage{mathrsfs}
\usepackage{amsthm}
\usepackage{setspace}
\usepackage{array}
\usepackage{float}
\usepackage{soul}
\usepackage{placeins}
\usepackage{mathtools}
\usepackage{braket}
\usepackage{xcolor}
\usepackage{graphicx}
\usepackage{epsfig}
\usepackage{pgfmath}
\usepackage{qcircuit}
\usepackage{savesym}
\usepackage{tabulary}
\usepackage{caption}
\usepackage{subcaption}
\usepackage{siunitx}

\usepackage{tabularx, booktabs}

\usepackage{titlesec}
\titleformat*{\section}{\large \bfseries}
\titleformat*{\subsection}{\normalsize \bfseries}
\titleformat*{\subsubsection}{\normalsize  \itshape}

\titlespacing*\section{0pt}{6pt plus 4pt minus 1pt}{1pt plus 2pt minus 2pt}
\titlespacing*\subsection{0pt}{6pt plus 4pt minus 1pt}{1pt plus 0pt minus 2pt}
\titlespacing*\subsubsection{0pt}{6pt plus 4pt minus 2pt}{1pt plus 2pt minus 2pt}

\author{Lilian Hunt}
\affiliation{%
  \institution{\it School of Computer Science \\University of Sydney}
  \city{}
  \country{}
}
\author{Alan Robertson}
\affiliation{%
  \institution{\it School of Computer Science \\University of Sydney}
  \city{}
  \country{}
}

\usepackage{etoolbox}% <-- for bold fonts
\newcommand{\ubold}{\fontseries{b}\selectfont}% <-- for bold fonts
\robustify\ubold% <-- for bold fonts
\newcommand{\ppercent}{\makebox[0pt][l]{\,\%}} % for percent symbol in table
\newcolumntype{C}[1]{>{\centering\let\newline\\\arraybackslash\hspace{0pt}}m{#1}}

\usepackage{physics} 
\usepackage{graphicx}

\usepackage{varwidth}
\usepackage{tikz}
\usetikzlibrary{decorations.pathreplacing,calligraphy}
\usetikzlibrary{fit}
\usetikzlibrary{arrows, arrows.meta, shapes.geometric, positioning, fit, arrows.meta, backgrounds, calc}
\usepackage{adjustbox}

\tikzset{%
  >={Latex[width=2mm,length=2mm]},
            base/.style = {rectangle, draw=black, text centered}
}
\tikzset{
    module/.style={%
        draw,
        inner sep=0.5em,
        },
    module/.default=2cm,
    >=LaTeX
}

\tikzset{fit margins/.style={/tikz/afit/.cd,#1,
    /tikz/.cd,
    inner xsep=\pgfkeysvalueof{/tikz/afit/left}+\pgfkeysvalueof{/tikz/afit/right},
    inner ysep=\pgfkeysvalueof{/tikz/afit/top}+\pgfkeysvalueof{/tikz/afit/bottom},
    xshift=-\pgfkeysvalueof{/tikz/afit/left}+\pgfkeysvalueof{/tikz/afit/right},
    yshift=-\pgfkeysvalueof{/tikz/afit/bottom}+\pgfkeysvalueof{/tikz/afit/top}},
    afit/.cd,left/.initial=2pt,right/.initial=2pt,bottom/.initial=2pt,top/.initial=2pt}

\newcommand{\todo}[1]{{\color{red} #1}}

\def\BibTeX{{\rm B\kern-.05em{\sc i\kern-.025em b}\kern-.08em
    T\kern-.1667em\lower.7ex\hbox{E}\kern-.125emX}}

\title{sQueeze: Accelerated Quantum Pulse Schedules}

\begin{document}
\begin{abstract}
\noindent \textit{Quantum devices in the Noisy Intermediate-Scale Quantum (NISQ) era are limited by high error rates and short decoherence times. Typically, compiler optimisations have provided solutions at the gate level. Alternatively, we exploit the finest level of quantum control and introduce a set of pulse level quantum compiler optimisations: sQueeze. Instead of relying on existing calibration that may be inaccurate, we provide a method for the live calibration of two new parameterised basis gates $R_{x}(\theta)$ and $R_{zx}(\theta)$ using an external server. We validate our techniques using the IBM quantum devices and the OpenPulse control interface over more than 8 billion shots. The $R_{x}(\theta)$ gates are on average 52.7\% more accurate than their current native Qiskit decompositions, while $R_{zx}(\theta)$ are 22.6\% more accurate on average. These more accurate pulses also provide up to a 4.1$\times$ speed-up for single-qubit operations and 3.1$\times$ speed-up for two-qubit gates. Then sQueeze demonstrates up to a 39.6\% improvement in the fidelity of quantum benchmark algorithms compared to conventional approaches.}
\end{abstract}

\date{}
\maketitle
\pagestyle{plain}

\section{Introduction}

\input{chapters/introduction.tex}
\section{Background}        
\label{section:background}
\input{chapters/background}

\section{Pulse Control of Transmon Qubits}
\input{chapters/related}

\section{sQueeze}        
\label{section:Model}
\input{chapters/model}

\section{Benchmarks}
\label{section:Methodology}
\input{chapters/methodology}

\section{Evaluation}  
\label{section:evaluation}
\input{chapters/evaluation}

\section{Conclusion}
\label{section:conclusion}
\input{chapters/conclusion}

\bibliographystyle{plain}
\bibliography{refs}

\clearpage

\appendix
\include{chapters/appendix}

\end{document}

%% file: chapters/introduction.tex
% end at one page 

% why do ppl care about quantum

% here are some problems
% - snap shot not representative
% - no live calibration 
% - no method to make it accessible

% answers

% why useful 3-5 contributions
% ideas for contributions:

Quantum computation may one day lead to essential applications in quantum chemistry~\cite{moll2018quantum}, finance~\cite{orus2019quantum, egger2020quantum}, optimisation~\cite{farhi2014quantum, egger2021warm}, and machine learning \cite{biamonte2017quantum, havlivcek2019supervised}. In contrast with the advantages promised by fault-tolerant quantum computation, modern `NISQ' devices are limited by high error rates, short decoherence times and low qubit counts~\cite{preskill2018quantum}. To realise the potential of quantum computing improvements are needed from the algorithmic level to the finest level of hardware control. For current transmon devices this control is expressed as sequences of parameterised `pulses'~\cite{alexander2020qiskit}.

% Quantum algorithms require arbitrary quantum operators, however devices only have a finite set of \textit{basis gates} with direct pulse implementations. 
While quantum algorithms require arbitrary precision operations, these operations must be implemented relative to a finite number of calibrated control operations for each quantum device. To achieve this, logical operations must be decomposed into a sequence of \textit{basis gates}, where only basis gates have a direct implementation on the device as a series of pulses. This decomposition to a sequence of basis gates is computationally non-trivial. Techniques have progressed from exhaustive searches~\cite{fowler2004constructing, kliuchnikov2015practical}, to limited depth searches~\cite{dawson2005solovay}, normal forms~\cite{bocharov2012resource, matsumoto2008representation}, and heuristic-based approaches~\cite{davis2020towards, mooney2021cost}. Pulse control started with brute-force optimization of a few pulse parameters \cite{jakubetz1990mechanism} and has advanced to Quantum Optimal Control (QOC)~\cite{de2011second, glaser2015training, krotov1995global}. However, most noisy experimental systems are not ready for QOC methods~\cite{gokhale2020optimized}. In addition to long compilation times, QOC depends on accurate models of the analogue pulses and the device Hamiltonian, which are difficult to determine experimentally and may fluctuate over short periods of time. 

Methods focusing on both gate and pulse optimisations have shown that adding parameterised operations at the pulse level reduce gate decomposition depth significantly~\cite{foxen2020demonstrating}. To avoid calibration costs, recent work has scaled the pulses calibrated by the device providers~\cite{earnest2021pulse, gokhale2020optimized}. However, our work has shown that the inferred pulse models are not necessarily a good approximation and rely on existing gate calibrations that may be inaccurate, especially for current noisy devices. Experiments conducted over five months show that current IBM quantum devices exhibit significant amounts of noise and that a single calibration snapshot is not necessarily a good representation of the average device behaviour.
\begin{figure}[t]
    \centering
    
    \begin{tikzpicture}
        % Nodes 
        \small
        \node (gate) {
        \Qcircuit @C=1em @R=.7em {
            & \ctrl{1} & \qw \\
            & \gate{\large R_x(\theta)} & \qw \\
            }
        };
        \node (gateLabel) [above =0.05em of gate] {Quantum Circuit};
        
        \node (qiskitLabel) [below = 0.5em of gate, xshift=-6.7em] {Standard Decomposition};
        
        \node (sQueezeLabel) [below = 0.5em of gate, xshift=6.7em] {sQueeze Decomposition};
        
        \node (decompQiskit)        [below=2.4em of gate, xshift=-6.7em]     {
                \Qcircuit @C=1em @R=.7em {
                & \qw & \ctrl{1} & \qw & \ctrl{1} & \qw & \qw \\
                & \gate{H} & \targ & \gate{R_z(\theta)} & \targ & \gate{H} & \qw 
            }
        };
        
        \node (decompSQueeze)        [below=1.9em of gate, xshift=6.7em]   {
                \Qcircuit @C=1em @R=.4em {
                    & \qw & \multigate{1}{R_{zx}(-\frac{\theta}{2})} & \qw \\
                    & \gate{R_x(\frac{\theta}{2})} & \ghost{R_{zx}(-\frac{\theta}{2})} & \qw \\
            }
        };
        
        \node (pulseSQueeze)    [below=7.2em of gate, xshift=6.7em]          {
            \includegraphics[width=0.23\textwidth]{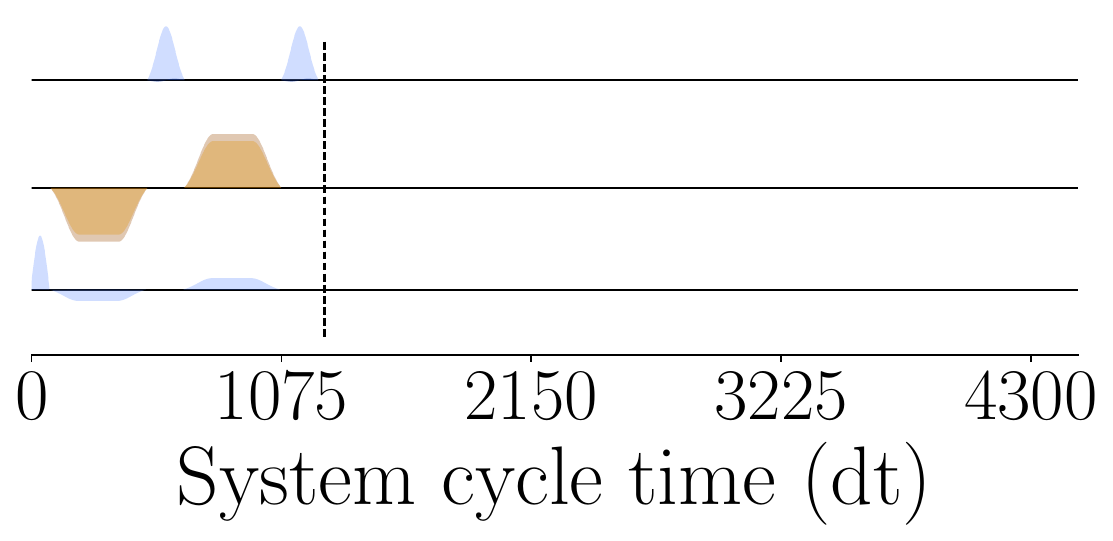} 
        };

        \node (pulseQiskit)    [below=7.2em of gate, xshift=-6.7em]          {
            \includegraphics[width=0.23\textwidth]{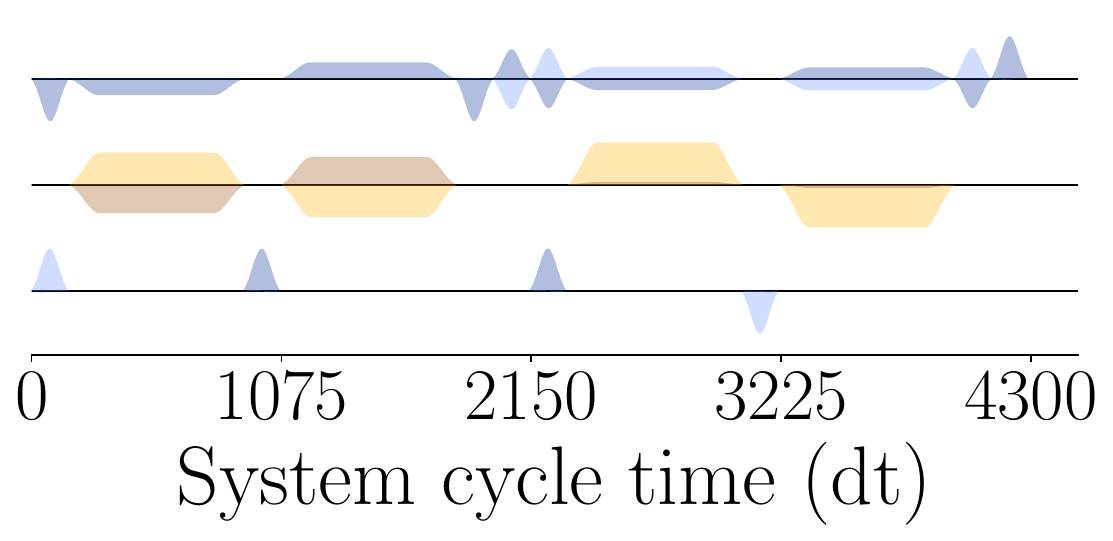}
        };
        
        % Arrows 
        \draw[->] (gate) -- (-2, -0.65) node [midway] {};
        \draw[->] (gate) -- (2, -0.65) node [midway] {};
        % \draw[->] (gate) -- (decompQiskit) node [midway] {};
        \draw[->] (decompSQueeze) -- ++(0, -1.15) node [midway] {};
        \draw[->] (decompQiskit) ++(0, -0.55) -- ++(0, -0.6) node [midway] {};
       
    \end{tikzpicture}
    \captionsetup{belowskip=0pt}
    \caption{sQueeze compilation pipeline in comparison to the native IBM implementation. The circuit is decomposed into a sequence of basis gates and then faster and higher fidelity pulses.}
    \label{fig:decompOverviewIntro}
\end{figure}
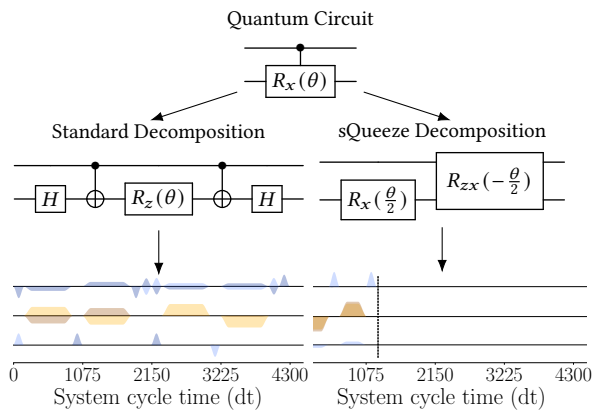

To join pulse-rescaling with better fidelity pulses we present {\it sQueeze}: a suite of software tools that perform both asynchronous pulse calibration and gate decompositions (see Fig.~\ref{fig:decompOverviewIntro}). We use 8 billion executions of the IBM quantum devices to extensively test sQueeze across a range of use cases to demonstrate the practicality of our work. Our contributions are as follows:
\begin{itemize}[leftmargin=1.5em, topsep=0.5pt]
    \item We demonstrate faster and more accurate parameterised $R_x(\theta)$ and $R_{zx}(\theta)$ gates with direct pulse constructions, we build these pulses using an asynchronous calibration server to track fluctuations in the backend quantum devices.
    \item We build a pass manager that decomposes a range of common quantum operations using the new set of basis gates.
    \item We perform extensive benchmarks of sQueeze compared to other state-of-the-art methods using tomography, NISQ circuit benchmarks, and randomised benchmarking.
    \item sQueeze demonstrates up to a 4.1$\times$ speed-up for SU(2) gates, 3.1$\times$ speed-up for SU(4) gates and up to a 39.6\% reduction in error for common quantum algorithms.
\end{itemize}

%% file: chapters/background.tex
We start with a brief introduction to quantum computing, focusing on aspects relevant to our work.
\subsection{Quantum Computation}
Quantum computation may be considered an extension of both probabilistic and reversible computing. The qubit is a unit of quantum memory that stores any unitary state described by $\ket{\psi} = \alpha\ket{0} + \beta\ket{1}$, where $\alpha$, $\beta \in \mathbb{C}$ and $|\alpha|^2 +|\beta|^2$ = 1. A more general multi-qubit system is described as a probability distribution over the convex set of these basis vectors; $\ket{\psi} = \sum^{\todo{2^n}}_i \alpha_i\ket{i}$ such that $\sum_i |\alpha_i|^2 = 1$. Quantum operations are rotations about the vector-space spanned by an ensemble of qubits. They are unitary and hence reversible operations that preserve the normalisation of the probability distribution over quantum states. As such, an arbitrary quantum operation $U$ can be represented as a matrix acting on a quantum state $\ket{\psi}$ by $U\ket{\psi_0} \rightarrow \ket{\psi_1}$. 

%  $\ket{0}$ and $\ket{1}$ are measurable computational basis states, which have $|\alpha|^2$ and $|\beta|^2$ probability of being observed, respectively.
% Operations are not legal if they do not preserve the normalisation of the probability distribution over quantum states, and as a result all quantum operations must be unitary. As unitary operators are linear we can represent an arbitrary quantum operation $U$ as a matrix acting on a quantum state $\ket{\psi}$ by $U\ket{\psi_0} \rightarrow \ket{\psi_1}$. As linear operators are invertible quantum operations in turn must be reversible.

\subsection{Compiling Quantum Circuits to Pulses}
Modern superconducting quantum systems are controlled via the capacitive coupling of a charge qubit and a microwave resonator~\cite{Qiskit, peterer2015coherence}. The physical evolution of a quantum system is described by a Hamiltonian operator $\mathscr{H}$. The relative evolution of the system depending on the state of the microwave resonator can then be considered a computational operation acting on that qubit. Computational operations on a qubit may then be achieved by driving the microwave resonator at a given amplitude for a given period of time. The action of this drive channel is termed a `pulse'. As the pulse control is itself generated by a classical computational system, the pulse control must be performed in discrete units of time.

% In an ideal noiseless world these pulses might be conceived of as a square pulse acting for a discrete unit of time with an associated linear response in the evolution of the system, however physical pulses have rise and fall periods. The total action of the Hamiltonian on the system must then include these fringes. The square pulse model may then be extended to include a time dependent function that describes the `shape' of the pulse based on  $f$.

The pulse model includes a time dependent function that describes the `shape' of the pulse based on $f$. The action on a quantum state $\psi_{t_0}$ from a control pulse $f$ acting from time $t_0$ to $t_1$ with parameters $\vec{\alpha}$ is then
\vspace{-3mm}
\begin{equation}
\ket{\psi_{t_1}} = e^{i\mathscr{H}\int_{t_0}^{t_1}f(t, \vec{\alpha})dt}\ket{\psi_{t_0}}
\vspace{-2mm}
\end{equation}
Two common parametric pulses are the {\it Derivative Removal by Adiabatic Gate} (DRAG) pulse for single-qubit operations~\cite{motzoi2009simple} and {\it Gaussian-Square} (GS) pulse for operations that couple pairs of qubits~\cite{sheldon2016procedure}. Techniques such as process and Hamiltonian tomography may be used to determine the set of control parameters that are required to implement a particular gate~\cite{Granade_2017}.
This abstraction from control pulses to computational operations allows a quantum program to neatly encapsulate all lower level implementation details within a gate and circuit model of computation. A sequence of computational gates is termed a {\it circuit}. A circuit can be represented as a diagram, for example in Fig.~\ref{fig:decompOverviewIntro}, where the lines represent qubits and the boxes represent operators acting on specific qubits. 

Quantum programs are specified at the gate level but to execute on a device the circuit must be mapped to sequences of control pulses for each qubit. The quantum compilation pipeline first decomposes operations into a sequence of \textit{basis gates} which are in turn decomposed into pulses and scheduled as seen in Fig.~\ref{fig:decompOverviewIntro}. The basis gates are the small set of one- and two-qubit operations that are implemented directly on the device as pulses. The basis gates are furthermore dependent on the qubit and quantum device; not all pairs of qubits support two-qubit operations. Typically the device \textit{coupling map} is sparse, which means swap operations must be inserted to ensure there are only operations between connected qubits~\cite{sabre}. 

\subsection{Tomography and Gate Calibration}
While driving a microwave resonator causes a quantum system to evolve, it does not provide any information as to what operation was applied to the quantum state. A common technique for diagnosing an unknown quantum operation is process tomography~\cite{merkel_tomography}. Process tomography is performed by preparing a set of quantum states in a spanning set of orthogonal bases, applying the unknown operation to each of them, and then measuring each of the outputs in a spanning set of orthogonal measurement bases~\cite{greenbaum_tomography}. By sweeping over a range of the drive parameters we can construct a quantum operation as a function of the pulse parameters~\cite{Granade_2017,quantum_process_tomography}. 
These raw pulses will typically perform a range of operations on the state. To construct a particular computational basis gate the pulse parameters must be tuned to remove unwanted components~\cite{howard}. In situations where this is not possible then a sequence of subsequent pulses may be constructed~\cite{sheldon2016procedure}. For well-understood physical systems the initial stage of process tomography may be replaced with Hamiltonian tomography~\cite{Granade_hamiltonian}.

\subsection{Gate Decomposition}
Having constructed a set of computational basis gates we are still left with an open problem: the set of all quantum operations is infinite but the basis set is strictly finite. Luckily, it has been shown that all multi-qubit gates can be decomposed to a set of universal single-qubit rotations and a single entangling two-qubit gate~\cite{barenco1995elementary}. As a result it is sufficient to calibrate one two-qubit gate (such as a CNOT) and a set of single-qubit gates to implement all rotations to an arbitrary accuracy. 

% These proofs provide a minimal guarantee that a gate may be constructed, but do not attempt to determine an efficient implementation in the number of basis gates. Solovay-Kitaev is simple approach to this problem, but this incurs a poly-logarithmic overhead in the implementation cost of each gate that is not in the basis set~\cite{dawson2005solovay}. This cost typically depends on the `distance' of the gate from the basis set as measured by iterative anti-commutation relations between elements of the basis set. As a result the overhead of implementing an arbitrary quantum operations may be reduced by increasing the number of calibrated basis gates.

A move towards parameterised gates is generally a more efficient approach for gate decomposition than search methods like Solovay-Kitaev~\cite{dawson2005solovay}. An arbitrary single-qubit operation may be described as a $U_3$ rotation~\cite{nielsenchuang2011}, which may in turn be decomposed into a sequence of $U_1$ rotations about each basis. By parameterising one of these $U_1$ gates (either $R_x(\theta), R_y(\phi)$ or $R_z(\gamma)$) along with a discrete set of operations that map between these bases, a single parameterised gate may then construct the full set of $U_3$ operations with constant overhead. The construction of larger arbitrary quantum operations can then be performed by a Weyl decomposition~\cite{khaneja2000cartan, zhang2004optimal}.

In the case of Qiskit~\cite{Qiskit}, the Hamiltonian of the system results in a $U_1$ gate in the $Z$ basis that acts linearly as a function of time. Parameterised $R_z(\gamma)$ gates may then be performed for free via software~\cite{mckay2017efficient}, resulting in the five gate $U_3$ decomposition:
\begin{equation}
    U_3(\theta,\phi,\lambda) = R_z(\phi+ 3\pi) \, \sqrt{X} \, R_z(\theta+\pi) \, \sqrt{X} \, R_z(\lambda).
\end{equation}
This requires two physical DRAG pulses instead of one because the $\sqrt{X}$ is not parameterised like the $R_z(\theta)$.
% In this work, we seek to extend the set of parameterised basis gates to include $R_x(\theta)$ along with parameterised two-qubit gates to improve circuit decompositions.

%% file: chapters/related.tex
% Related 
Now, we explore some existing pulse optimisation techniques that use models of the pulse area to add parameterised gates to the basis set. These methods can improve gate fidelity and schedule duration without requiring further calibration.

\subsection{Single-Qubit Pulse Scaling}
The amplitude of Qiskit's $X$ pulse can be scaled by $\frac{\theta}{\pi}$ to achieve $R_x(\theta)$ rotations~\cite{gokhale2020optimized}. Now single-qubit gates can be implemented with only one direct $R_x(\theta)$ gate as follows,
\begin{equation} \label{eq:gokhaleu3}
    U_3(\theta,\phi,\lambda) = R_z(\phi + \pi) \, R_x(\theta)\, R_z(\lambda-\pi).
\end{equation} 
This halves the schedule duration for an arbitrary $U_3$ gate from $320\, \mathtt{dt}$ to $160\mathtt{dt}$, where $\mathtt{dt}$ is Qiskit’s standard unit for duration and $\mathtt{dt} = 0.22ns$ \cite{gokhale2020optimized}. The $1.6\times$ improvement in fidelity is attributed to three sources~\cite{gokhale2020optimized}: shorter pulses, less calibration error susceptibility, and smaller pulse amplitudes.

\subsection{Two-Qubit Pulse Scaling}
\begin{figure}[ht]
    \centering
    \includegraphics[width=\linewidth]{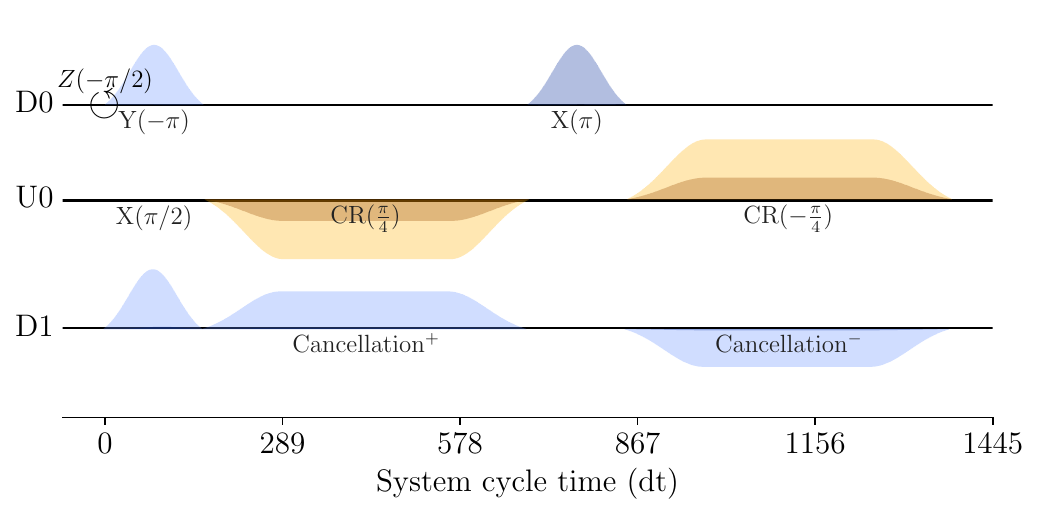}
    \caption{Qiskit CNOT pulse schedule where D0 and D1 are the drive channels, and the control channel U0 couples the qubits. The cancellation pulses on D1 compensate for unwanted terms.}
    \label{fig:cnotSchedule}
\end{figure}

Qiskit only calibrates one two-qubit basis gate~\cite{Qiskit}: a CNOT. This results in longer decompositions than if a parameterised two-qubit gate was part of the basis set~\cite{foxen2020demonstrating}. Sometimes this CNOT is built from echoed cross-resonance $CR\bigl(\pm \frac{\pi}{4}\bigr)$ pulses \cite{chow2011simple, sheldon2016procedure} with cancellation pulses on the target qubit \cite{sundaresan2020reducing} and single-qubit pulses as shown in Fig.~\ref{fig:cnotSchedule}. The $CR\bigl(\pm \frac{\pi}{4}\bigr)$ are Gaussian-Square pulses with a flat-top amplitude $A$, width $w$, and Gaussian flanks with standard deviation $\sigma$. Their area is $F = |A|[w + \sqrt{2\pi}\sigma \text{erf}(n_\sigma)]$~\cite{earnest2021pulse}. 
% \begin{figure}[h]
%     \centering
%     \scalebox{0.9}{
%         \Qcircuit @C=1em @R=1.2em  @!R {
%             & \gate{R_z\bigl(\frac{\pi}{2}\bigr)} & \gate{R_y(-\pi)} & \multigate{1}{R_{zx}\bigl(\frac{\theta}{2}\bigr)} & \gate{X} & \multigate{1}{R_{zx}\bigl(- \frac{\theta}{2}\bigr)} & \gate{X} & \gate{X} & \qw \\
%             & \gate{R_x\bigl(\frac{\pi}{2}\bigr)} & \qw & \ghost{R_{zx}\bigl(\frac{\theta}{2}\bigr)} & \qw & \ghost{R_{zx}\bigl(- \frac{\theta}{2}\bigr)} & \qw & \qw & \qw  \gategroup{1}{4}{2}{7}{.8em}{--}
%     }}
%     \captionsetup{aboveskip=1em}
%     \caption{Qiskit CNOT pulse schedule, the dashed box represents the $R_{zx}$ pulse schedule and last two $X$ gates annihilate.}
%     \label{fig:cnotPulseScheudle}
% \end{figure} 

In this case, IBM's calibrated cross-resonance pulses can be scaled to provide direct $R_{zx}(\theta)$ without further device experiments. To implement arbitrary $CR(\theta)$ rotations the Gaussian-Square pulses are scaled in respect to 2$\theta F /\pi$, where $F$ is the original pulse area. Then $R_{zx}(\theta)$ can be implemented as $R_{zx}(\theta) = CR(\frac{\theta}{2})\, X\, CR(-\frac{\theta}{2})\, X$ where $X$ is applied to the control channel. Adding the new parameterised $R_{zx}(\theta)$ gates to the basis set and using Cartan's decomposition has been shown to improve the fidelity by up to 38\% and yield a 52\% schedule 
time reduction for arbitrary SU(4) decompositions \cite{earnest2021pulse, gokhale2020optimized}.

% To create a target area $\alpha(\theta)$ they first scale $w$, then when $\alpha(\theta) < |A| \sigma \sqrt{2\pi} \text{erf}(n_\sigma )$ they set $w = 0$ and scale the pulse amplitude such that $|A(\theta )| = \alpha(\theta)/[\sigma \sqrt{2\pi} \,  \text{erf}(n_{\sigma})]$.

%% file: chapters/model.tex
% \todo{what problems in previous techniques are we looking to work from, these are the problems, here's what the model does, here's how we solve them }
From the previous methods we have two main challenges: extending the set of parameterised basis gates to enable speed-ups, and ensuring the accuracy of computation does not suffer.

To achieve these goals we propose sQueeze; a combined calibration server and pulse compilation pass. The calibration server provides live calibrations for direct $R_x(\theta)$ and $R_{zx}(\theta)$ pulses. The compilation pass then decomposes logical quantum operations in terms of this extended set of pulses that use the scaling parameters provided by the calibration server. This two stage asynchronous approach ensures that the user is able to compile their circuit with the most up-to-date pulse parameters without being blocked by device calibrations.

% The previous methods have shown that scaling single-qubit DRAG and two-qubit Gaussian-Square pulses is feasible. However, they only explore downscaling the pulses to implement smaller rotations and do not explore finding faster initial pulses. There has been no existing work on scaling both the width and amplitude of DRAG pulses. These methods also rely on existing calibration that may not be accurate, instead we provide new techniques for tracking device performance and live calibration.

% However, Rabi sweeps over the possible amplitudes and durations show it is possible to implement the full range of $R_x(\theta)$. 
% The optimal duration is dependent on the device and qubit, a list of optimal duration for all benchmark devices can be found in Appendix \ref{appendix:qubitdurations}.

\subsection{sQueeze $R_{x}(\theta)$ Pulses}\label{model:1qSection}
\begin{figure}
    \centering
    \includegraphics[width=\linewidth]{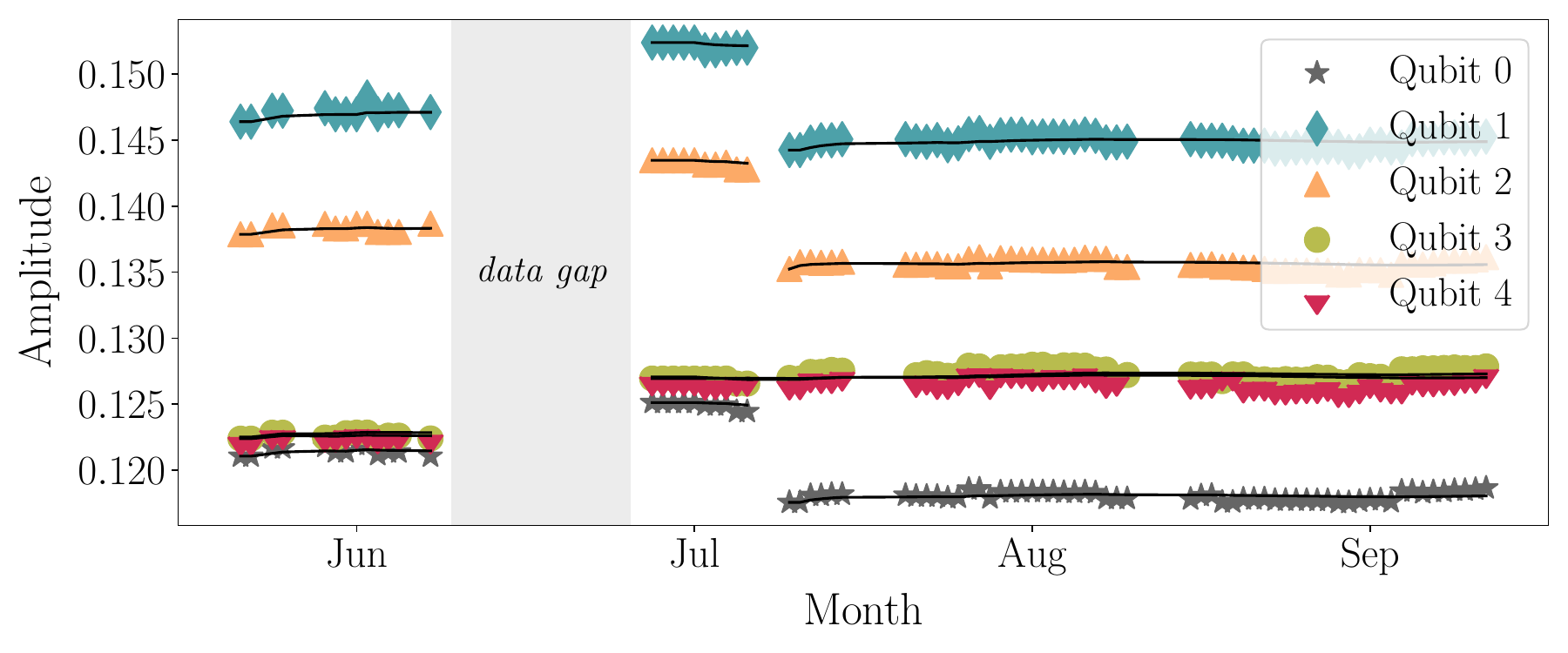}
    \small
    \caption{Amplitude of Qiskit's $X$ pulse over four months for IBM Lima. There is a data gap between days 20 and 37.}
    \label{fig:lima_cal_amp_over_time}
\end{figure}

We extend the existing DRAG pulse scaling to reshape the duration in addition to the amplitude and implement $R_x(\theta)$ rotations faster than $160\, \mathtt{dt}$~\cite{gokhale2020optimized}. sQueeze begins by performing a sweep over $\text{DRAG}(A, t, \sigma, \beta)$ pulses with parameters $64\, \mathtt{dt} \le t \le 160\, \mathtt{dt}$, $0 \leq A \leq 1$ in order to find the fastest $R_x(\pi)$ operation. Note, $\sigma, \beta$ are the same as Qiskit's $X$ pulse parameters. This fastest $X$ gate is then parameterised by $A = A_0$ and $t = t_0$. After finding this fastest $X$ operation, calibration circuits are run over the range $0 \le A \le A_0$ and the data is fit to an ideal $\text{sin}^2\bigl(\frac{\theta}{2}\bigr)$ function using the Levenberg-Marquardt curve-fitting algorithm~\cite{levenberg1944method}. To calculate the amplitude required to implement any $R_x(\theta)$ gate, we equate the fitted $A_1 \, \text{sin}^2(\omega A + \phi) + \delta$ function to the ideal probability distribution function to give,
\begin{equation}\label{eq:sin2curve}
\setlength{\abovedisplayskip}{2pt}
\setlength{\belowdisplayskip}{1pt}
     A_1 \, \text{sin}^2(\omega A + \phi) + \delta = \text{sin}^2\Bigl(\frac{\theta}{2}\Bigr) 
\end{equation} 
\begin{equation}
\setlength{\abovedisplayskip}{0pt}
\setlength{\belowdisplayskip}{0pt}
     A = \frac{\text{sin}^{-1}\Biggl(\sqrt{\frac{\text{sin}^2\Bigl(\frac{\theta}{2}\Bigr) - \delta}{A_1}}\Biggr) - \phi}{\omega}
\end{equation}
$A$ is then the applied amplitude of the DRAG pulse for $R_x(\theta)$. As the single-qubit pulses are stable over the course of days to weeks, seen in Fig~\ref{fig:lima_cal_amp_over_time}, we run the calibration procedure on average once every two hours and use trailing averages to determine the $\sin^2$ parameters.
 % Over four months the amplitude of Qiskit's $X$ gate only had a standard deviation 1.2\%. Over short periods of time, the amplitude drifts around a given mean, for example between days 1 to 20 in Fig.~\ref{fig:lima_cal_amp_over_time}. However, over longer periods of time, the system may shift and we observe jumps in the mean calibrated amplitude for each qubit, as seen on day on 50 in Fig.~\ref{fig:lima_cal_amp_over_time}. The changes in the amplitude of Qiskit's $X$ pulse are reflected in changes to the fitted frequency $\omega$ in Equation \ref{eq:sin2curve}. For example, when the amplitude of $X$ pulse for qubit 0 drops from 0.125 to 0.118 on day 50, $\omega = 3.37$ becomes $\omega = 3.46$, which in turn changes $A$ the required amplitude to drive a $\theta$ rotation about the $\hat{x}$ axis.

\begin{figure}
    \centering
    \includegraphics[width=\linewidth]{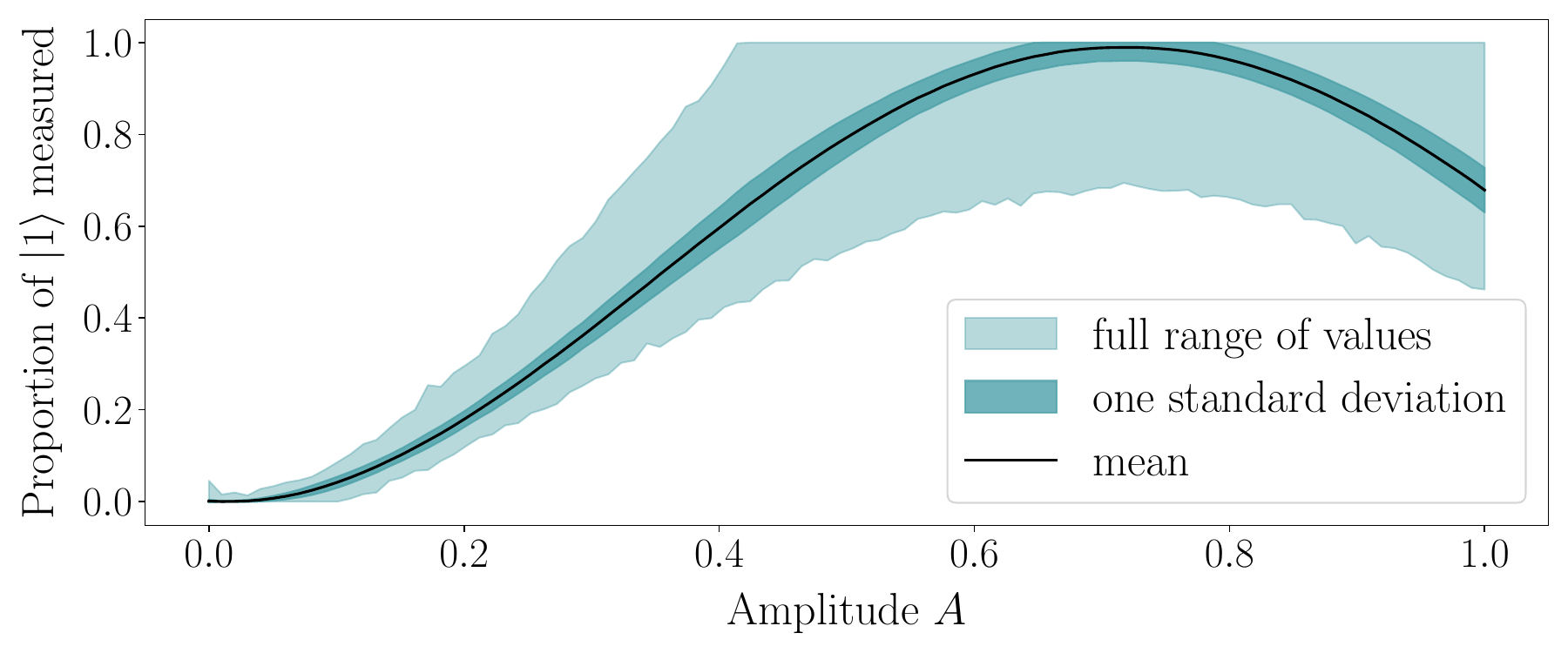}
    \small
    \caption{IBM Nairobi Qubit 6 testing $\text{DRAG}(A, d, \sigma, \beta)$ pulses for $0 \leq A \leq 1$ over $2.6 \times 10^{6}$ shots and 4 months.}
    \label{fig:nairobiSinData}
\end{figure}

While the pulses are stable over the medium to long term, IBM devices experience significant short term fluctuations. Fig.~\ref{fig:nairobiSinData} shows the proportion of $\ket{1}$ states measured over four months and 260 million shots. The light blue shaded region highlights the full range of results that show points with error up to 40\%. However, the darker region represents one standard deviation and shows that the experimental data matches the ideal $\sin^2\bigl(\frac{\theta}{2}\bigr)$ curve on average. Hence, we use the mean of the periodic calibration data over a two day window to reduce the influence of short term noise on the system.

% Consequently, we only use the measurement data from the last two days to fit the $\sin^2$ curve. This balances using the most recent data and ensuring there are enough data points such that outliers do not significantly skew the data. To handle sudden device shifts, we only include data collected when the amplitude of Qiskit's $X$ pulse is within 0.001 of the current calibrated amplitude. 
% This trailing average is drawn in black for each qubit in Fig.~\ref{fig:lima_cal_amp_over_time}. Now, we can construct a new faster and more efficient DRAG($A$ $d$, $\sigma$, $\beta$) pulses.

\subsection{sQueeze $R_{zx}(\theta)$ Pulses}\label{model:2qSection}
sQueeze extends the Earnest CR scaling method to find faster CNOT gates. After the initial Earnest scaling, we use a local amplitude sweep to fine tune the resulting CR pulse accuracy. Limited by computational power and device queues, we use particle filtering to search for the scaling factors. Then we use the original Earnest method to implement faster $R_{zx}(\theta)$ gates. 

\subsubsection{CR Pulse Reshaping}
\textit{ }We find that performing a na\"ive rescaling of Qiskit's initial pulse envelope $GS(A, w, d, \sigma)$ to halve the duration of the CR pulse~\cite{earnest2021pulse} increases the average CNOT error rate from $2.33\%$ to $6.18\%$. This test was performed using $26.4$ million shots over all qubits on IBM Quito, Manilla, Belem and Lima. Due to this increased error rate we supplement the initial pulse-rescaling with a local parameter search and reshape the pulses again in order to improve the accuracy of the scaled CR pulse. When reshaping these pulses we are constrained by the discrete intervals of $16\, \mathtt{dt}$ over which pulses are defined, while the normalised amplitude $A$ is a continuous complex number such that $0 < |A| \leq 1$. 

\begin{figure}[h]
    \centering
    \includegraphics[width=\linewidth]{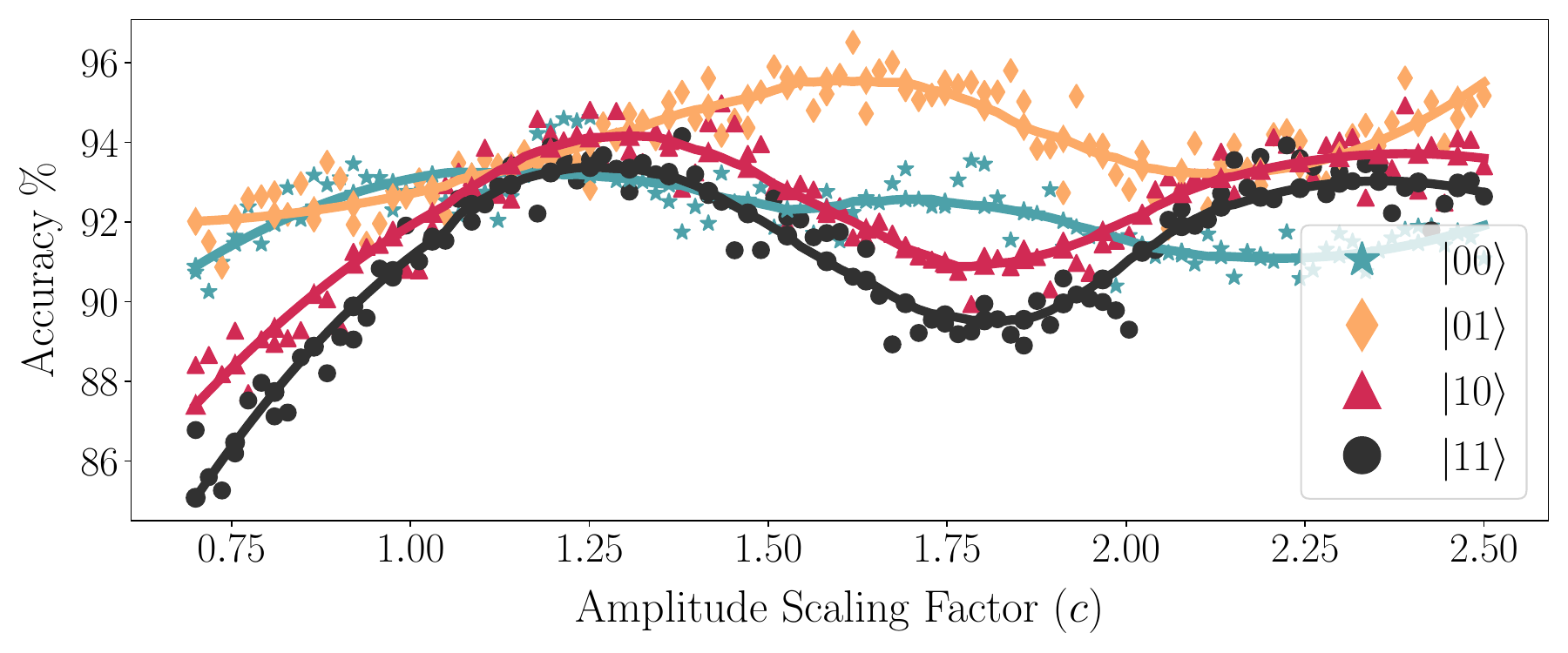}
    \caption{IBM Nairobi CNOT (0, 1) accuracy for the $Z$ basis as the scaling factor $c$ varies over 1.6 million shots.}
    \label{fig:cnotAmpFudgeStartingStates}
\end{figure}
\begin{figure}
    \centering
    \begin{subfigure}[b]{0.48\linewidth}
        \centering
        \includegraphics[width=\textwidth]{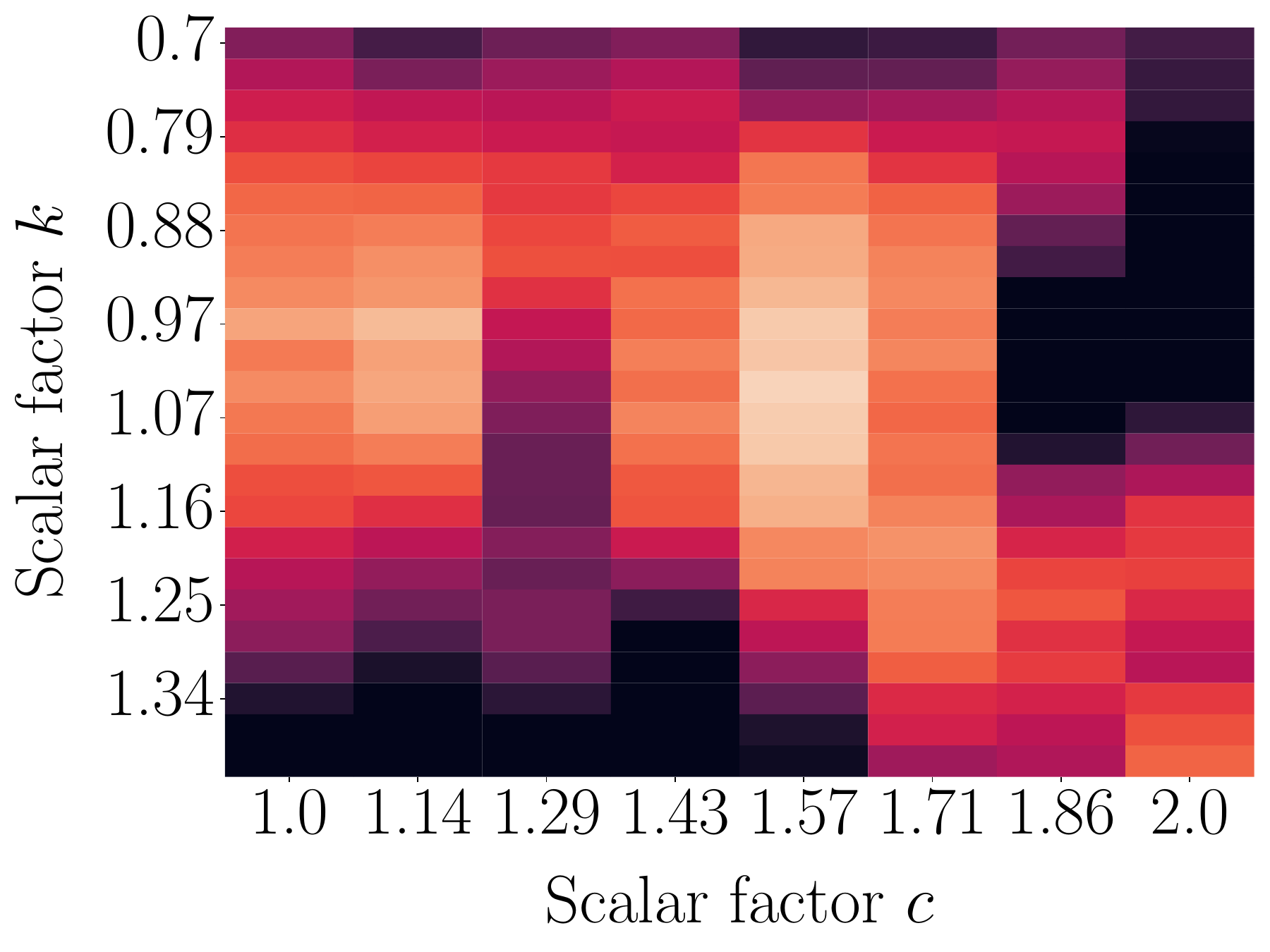}
        \captionsetup{justification=raggedright,margin={3.7em, 0em},singlelinecheck=off}
        \caption{\footnotesize IBM Lima (1, 3)}
        \label{fig:cnotHeatmapLima13}
    \end{subfigure}
    \begin{subfigure}[b]{0.49\linewidth}  
        \centering 
        \includegraphics[width=\textwidth]{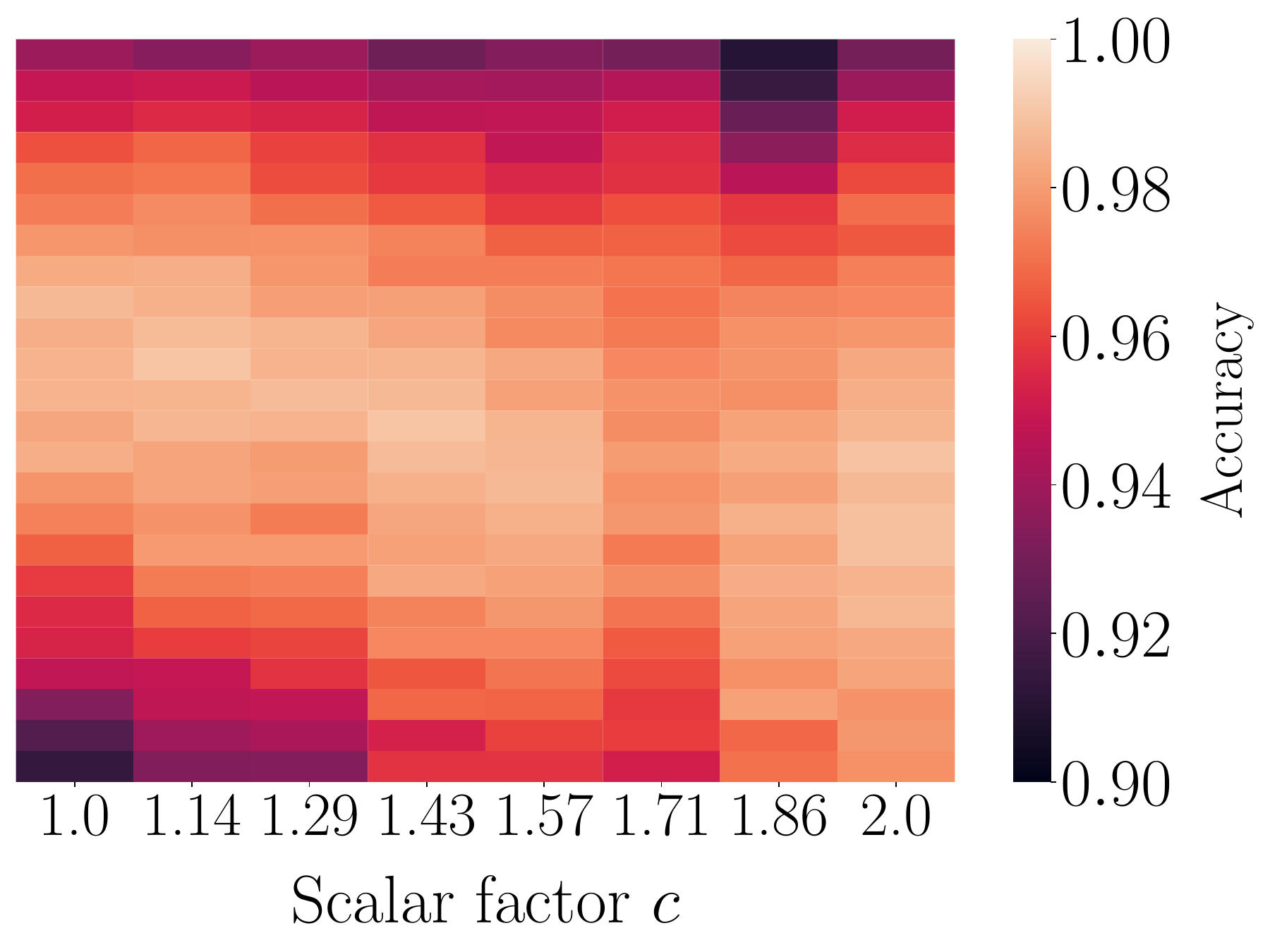}
        \captionsetup{justification=raggedright,margin={1.2em, 0em},singlelinecheck=off}
        \caption{IBM Belem (3, 4)}
        \label{fig:cnotHeatmapBelem34}
    \end{subfigure}
    \caption{Example CNOT accuracy averaged over the four basis states and 3.1 million shots per qubit pair as $k$ and $c$ vary. The heatmap is unique for each device and qubit pair.}
    \label{fig:cnotHeatmaps}
\end{figure}


sQueeze begins rescaling the pulse by increasing the amplitude and decreasing the pulse duration by a scalar factor $c$.  Our initially reshaped pulse is described by a rescaled amplitude $A' = cA$ width $w' = \frac{F - \sigma \norm{A'} \sqrt{2\pi}}{A'}$, a rescaled duration $d' = 16 \left\lfloor\frac{w' + \sigma n_\sigma}{16}\right\rfloor$ where $n_\sigma = \frac{d_0 - w_0}{\sigma}$ and $F$ is the area of the original pulse envelope. To improve the accuracy of this reshaped pulse we sweep over nearby amplitudes using a fine tuning factor $k$. The new CR pulse is built using Gaussian-Square pulses $GS(kA', w', d', \sigma)$. 
The need for this $k$-fine-tuning can be seen in Fig.~\ref{fig:cnotAmpFudgeStartingStates}, where $c$-scaled CR pulses were used to construct CNOT operations and the average accuracy of each computational basis state varies as the scalar factor $c$ changes. The results of the $c$-scaling and the $k$-fine-tuning sweep for CR pulses on two different devices can be seen in Fig~\ref{fig:cnotHeatmaps}. For each choice of $c$ there is a choice of $k$ that improves the accuracy of the constructed gate.

% To determine the fitness of a particular choice of $c$ and $k$ we construct a CNOT using the $CR$ pulses test the application of this gate to the $Z$ basis states $\ket{00}, \ket{01}, \ket{10}$, and $\ket{11}$. 

% This is highlighted in Fig.~\ref{fig:cnotAmpFudgeStartingStates} where the $\ket{11}$ accuracy is lowest around $k = 1.75$ but the $\ket{01}$ state is most accurate here. We explore the space by performing a grid search of combinations of $n_c$ equally spaced values of $c$ ranging from $c_{\min}$ to $c_{\max}$, and $n_k$ equally spaced values $k$ ranging from $k_{\min}$ to $k_{\max}$. Not only is the CNOT performance state dependent, the distribution of accurate points also depends on the pair of qubits and device. We can see that the heatmaps are very different for IBM Lima Qubits (1, 3) and IBM Belem Qubits (3, 4) in Fig.~\ref{fig:cnotHeatmapLima13} and Fig.~\ref{fig:cnotHeatmapBelem34}, respectively.

\subsubsection{Pulse Parameter Particle Filtering}
\textit{ }A full grid search is informative but extremely resource intensive. Testing  $n_c = 24$ and $n_k = 8$ scalar factors in the four basis states as shown in Fig.~\ref{fig:cnotHeatmaps}, requires 8 batch requests per qubit pair. This takes on average 24 hours, given IBM's public queue times. 

Instead of executing a full grid search, we build off the historical grid data and perform particle filtering: an alternative search method that does not require a differentiable or continuous function~\cite{djuric2003particle}. We begin particle filtering by defining a region in the $c-k$-domain that has historically produced relatively good results for that particular qubit pair. For an initial pass we regularly space a grid of particles $p_{0,i} = (c_{0, i}, k_{0, i})$ over this region, along with a particle at $c=1$, $k=1$ to implement the default Qiskit strategy as a baseline. For each particle we construct a CNOT operation and apply it against four computational states $\ket{00}, \ket{01}, \ket{10},$ and $\ket{11}$. The resulting accuracy of the particle then defines an associated weight $\alpha_{0,i}$. The distribution of particle weights now approximates the accuracy function over the $c-k$ domain. As these distributions are relatively flat we increase the gradient by raising each score to the power of $32$ before the weights are normalised.

\begin{figure}
    \centering
    \begin{subfigure}[b]{0.49\linewidth}
        \centering
        \includegraphics[width=\textwidth]{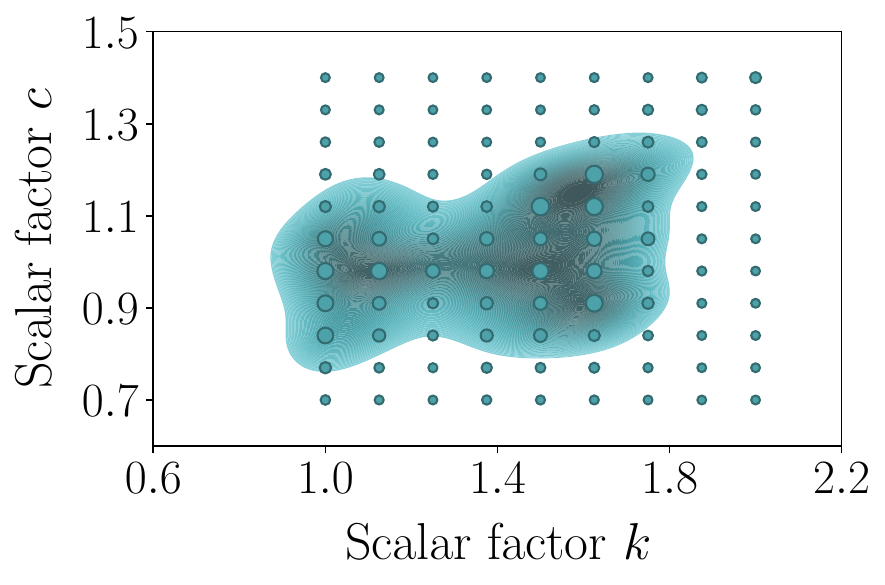}
        \captionsetup{justification=raggedright,margin={4em, 0em},singlelinecheck=off}
        \caption{\footnotesize Iteration 1}
        \label{fig:particle_iter1}
    \end{subfigure}
    \hfill
    \begin{subfigure}[b]{0.49\linewidth}  
        \centering 
        \includegraphics[width=\textwidth]{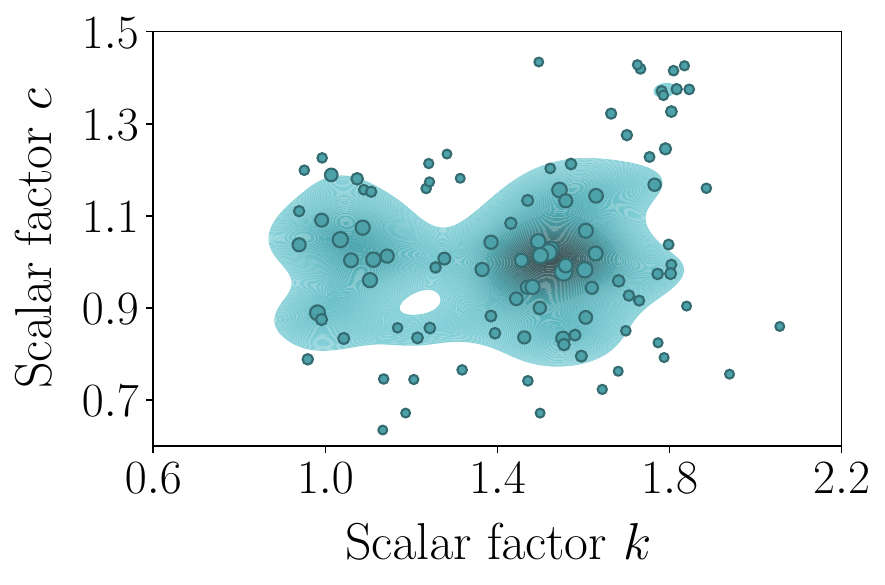}
        \captionsetup{justification=raggedright,margin={4em, 0em},singlelinecheck=off}
        \caption{\footnotesize Iteration 2}
        \label{fig:particle_iter2}
    \end{subfigure}
    \begin{subfigure}[b]{0.49\linewidth}   
        \centering 
        \includegraphics[width=\textwidth]{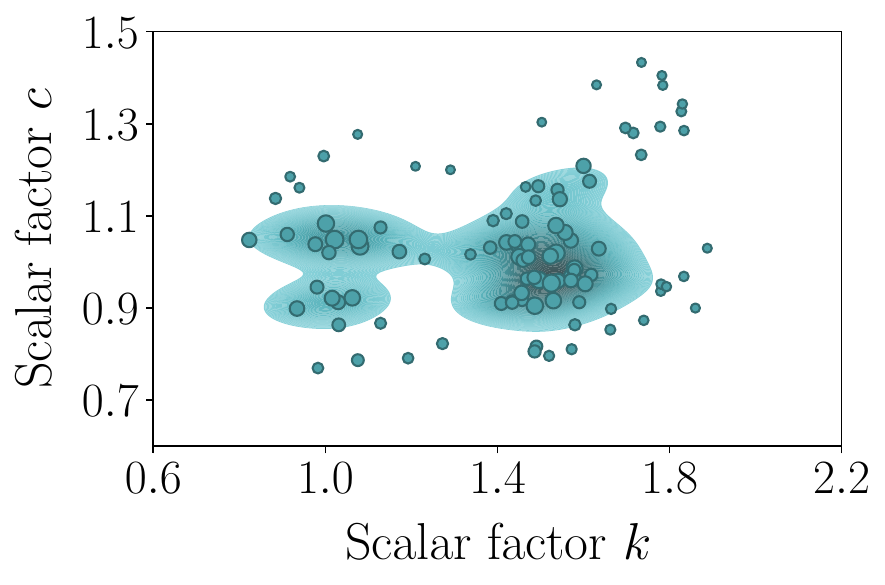}
        \captionsetup{justification=raggedright,margin={4em, 0em},singlelinecheck=off}
        \caption{\footnotesize Iteration 3}
        \label{fig:particle_iter3}
    \end{subfigure}
    \hfill
    \begin{subfigure}[b]{0.49\linewidth}   
        \centering 
        \includegraphics[width=\textwidth]{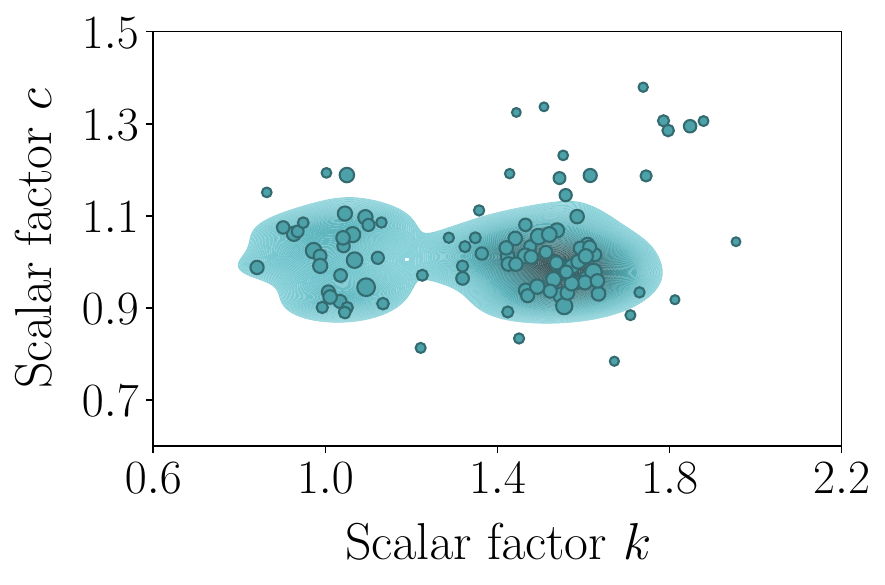}
        \captionsetup{justification=raggedright,margin={4em, 0em},singlelinecheck=off}
        \caption{\footnotesize Iteration 4}
        \label{fig:particle_iter4}
    \end{subfigure}
    \caption{Example of particle filtering for IBM Lima Qubits (1, 3). Point size is proportional to $a^{p}$, $a$ is accuracy and $p=32$.}
    \label{fig:particleFiltering4Iterations}
\end{figure}

To narrow our domain we distribute a new selection of particles based on the function approximated by the particle weights of the previous round. This encourages the clustering of points around known good selections of pulses. New particles $\vec{p}_{n}$ are drawn from the previous round's particles such that $p_{n-1,i}$ is selected with probability $\alpha_{n-1, i}$. Each of these re-sampled particles is perturbed by a small amount of Gaussian noise. The standard deviation of this noise is set to half the size of the initial particle spacing. At this point we add two special particles; the best particle from the previous round and Qiskit's particle. The most accurate particle from the previous round is the one with the highest weight, if there is a tie then we choose the particle with the largest $c_i$ as we prefer faster pulses. This set $\vec{p}_{n}$ now represents the next round of input for the particle filtering method. An illustrative example is shown in Fig.~\ref{fig:particleFiltering4Iterations}. By the fourth iteration seen in Fig.~\ref{fig:particle_iter4}, the points form two clusters that match the most accurate regions of the original grid shown in Fig.~\ref{fig:cnotHeatmapLima13}.

Furthermore, we only test particles with $c_i \geq 1$ to ensure both a faster and more accurate pulse. As Qiskit's CR parameters are included in each round, sQueeze will always perform at least as well as Qiskit. If the best non-Qiskit particle is worse than Qiskit then the process resets the particles to the initial grid. This reset occurs on average once every 43 rounds of particle filtering. This process is performed asynchronously by the calibration server, which reports the current best selection of particles. A new round of particle filtering is performed on average once every two hours. These scaled cross-resonance pulses can now be used to define improved $R_{zx}(\theta)$ pulses using the Earnest method outlined in Sec.~\ref{section:background}. 

\subsection{Optimising Decompositions} \label{model:equivRelationSection}

The above strategies proposed extend the set of basis gates with direct pulse implementations but this does not immediately translate into more efficient decompositions for more complex gates. We must also provide decompositions in terms of our newly parameterised $R_x(\theta)$ and $R_{zx}(\theta)$.

\subsubsection{Single-Qubit Gate Decompositions} \label{equiv:singleQubit}
\textit{ }As $R_z$ operations are `free' operations that are considered to be perfectly accurate on IBM devices~\cite{mckay2017efficient}, all single-qubit rotations are benchmarked in terms of the duration of $R_x$ gates. We have previously demonstrated a direct $R_x(\theta)$ gate for $0 \le \theta \le \pi$, but with the addition of two free $R_z$ operations we implement negative rotations with the same schedule time and accuracy as our direct $R_x(\theta)$ by
\begin{equation}
    R_x(-\theta) = R_z(\pi) \, R_x(\theta) \, R_z(\pi).
\end{equation}
 % \ref{model:1qSection}.

\begin{figure}[ht]
    \centering
    \includegraphics[width=\linewidth]{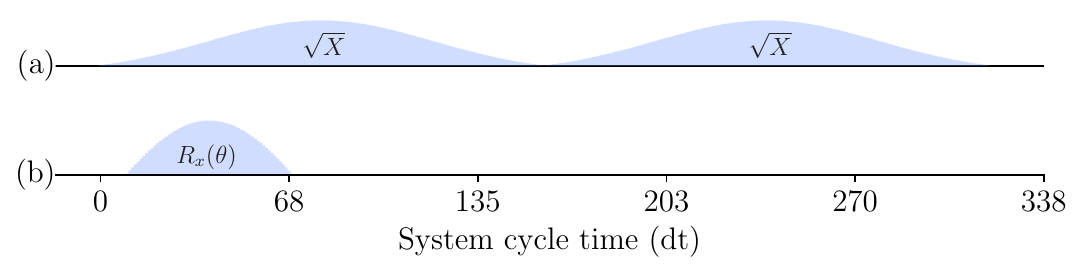}
    \caption{Standard compilation $U_3$ pulse schedule (a) vs sQueeze (b). sQueeze takes $78.12\, \mathtt{dt}$ on average, more than four times faster than Qiskit's $320\, \mathtt{dt}$ schedule where $1\mathtt{dt} = 0.22ns$.}
    \label{fig:pulseShapingDRAG}
\end{figure}

Equipped with a parameterised $R_x(\theta)$, free $R_z$ operations and Eq.~\ref{eq:gokhaleu3} we find a decomposition of a $U_3$ gate by
\begin{equation} \label{eq:u3zxzdecomp}
    U_3(\theta,\phi,\lambda) = R_z\Bigl(\phi +\bigl(\frac{\pi}{2}\bigr)\Bigr)  \, R_x(\theta) \, R_z\Bigl(\lambda-\bigl(\frac{\pi}{2}\bigr)\Bigr).
\end{equation}
The total pulse schedule length and accuracy now only depend on a single fast $R_x$ pulse instead of two standard $\sqrt{X}$ pulses, as illustrated in Fig.~\ref{fig:pulseShapingDRAG}.

\subsubsection{Multi-Qubit Gates}
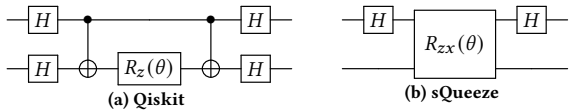
\begin{figure}[ht]
    \centering
    \begin{subfigure}[t]{0.47\linewidth}
        \centering
        \scalebox{0.9}{
        \Qcircuit @C=1em @R=.9em {
            & \gate{H} & \ctrl{1} & \qw & \ctrl{1} & \gate{H} & \qw \\
            & \gate{H} & \targ & \gate{R_z(\theta)} & \targ & \gate{H} & \qw 
            }
        }
        \captionsetup{width=0.95\textwidth}
        \caption{\footnotesize Qiskit}
        \label{fig:qiskitRxx}
    \end{subfigure}
    \begin{subfigure}[t]{0.47\linewidth}
        \centering
        \scalebox{0.9}{
        \Qcircuit @C=1em @R=1.2em {
            & \gate{H} & \multigate{1}{R_{zx}(\theta)} & \gate{H} & \qw \\
            & \qw & \ghost{R_{zx}(\theta)} & \qw  & \qw \\
            }
        }
        \captionsetup{width=0.95\textwidth}
        \caption{\footnotesize sQueeze}
        \label{fig:weylRxx}
    \end{subfigure}
    \caption{Comparison of Qiskit's $R_{xx}(\theta)$ decomposition into multiple CNOT gates compared to sQueeze's Weyl decomposition using a direct $R_{zx}(\theta)$ as a basis gate. Each Qiskit CNOT is further decomposed into more pulses as seen in Fig.~\ref{fig:cnotSchedule}.}
    \label{fig:rxxDecomp}
\end{figure}

\textit{ }For larger gates we use the Weyl chamber decomposition~\cite{zhang2003geometric, zhang2004optimal} with $R_{zx}$ as a basis gate. $R_{xx}$, $R_{yy}$ and $R_{zz}$ can all be implemented using a single $R_{zx}$ pulse, halving the number of two-qubit operations, compared to decompositions using CNOTs. An example of the Qiskit's $R_{xx}$ decomposition compared to the Weyl decomposition is shown in Fig.~\ref{fig:rxxDecomp}. Two-qubit gates can now be implemented up to 50\% faster~\cite{earnest2021pulse}. In particular, $C(\sqrt{X})$, an important gate for the Toffoli decompositions below, can now be decomposed into a single $R_x(\frac{\theta}{2})$, $R_{zx}(-\frac{\theta}{2})$, and a free $R_z(\frac{\pi}{4})$.

\subsubsection{Toffoli Decomposition}\label{model:Toffoli}
\begin{figure}[h]
    \begin{subfigure}{\linewidth}
        \centering
        \scalebox{.80}{
            \Qcircuit @C=.8em @R=0.5em @!R {
            & \qw & \qw & \qw & \ctrl{2} & \qw & \qw & \qw & \ctrl{2} & \ctrl{1} & \gate{T} & \ctrl{1} & \qw  \\
            & \qw & \ctrl{1} & \qw & \qw & \qw & \ctrl{1} & \gate{T} & \qw & \targ & \gate{T^\dag} & \targ & \qw \\
            & \gate{H} & \targ & \gate{T^\dag} & \targ & \gate{T} & \targ & \gate{T^\dag} & \targ & \gate{T} & \gate{H} & \qw & \qw   \\
            }
        }
        \caption{\footnotesize Default Qiskit decomposition using only single-qubit gates and CNOTs.}
        \label{fig:qiskitToffoli}
    \end{subfigure}
    \begin{subfigure}{\linewidth}
    \centering
        \scalebox{.80}{
            \Qcircuit @C=.8em @R=0.5em @!R {
            & \qw & \ctrl{1} & \qw & \qw & \qw & \ctrl{1} & \ctrl{2} & \qw \\
            & \ctrl{1} & \targ & \gate{S^\dag}  & \ctrl{1} & \qw & \targ & \qw & \qw \\
            & \gate{\sqrt{X}} & \gate{Z} & \qw & \gate{\sqrt{X}} & \gate{Z} & \qw & \gate{\sqrt{X}} & \qw 
            \gategroup{1}{7}{3}{8}{1em}{--}
        }
        }
        \caption{\footnotesize sQueeze decomposition using $C(\sqrt{X})$, the last two gates commute.}
        \label{fig:myToff1}
    \end{subfigure}
    % \begin{subfigure}{\linewidth}
    %     \centering
    %     \scalebox{.90}{
    %         \Qcircuit @C=.8em @R=0.5em @!R {
    %         & \qw & \ctrl{1} & \qw & \qw & \qw & \ctrl{2} & \ctrl{1} & \qw \\
    %         & \ctrl{1} & \targ & \gate{S^\dag} & \ctrl{1} & \qw & \qw & \targ & \qw \\
    %         & \gate{\sqrt{X}} & \gate{Z} & \qw & \gate{\sqrt{X}} & \gate{Z} & \gate{\sqrt{X}} & \qw & \qw  \\
    %         }
    %     }
    %     \caption{\footnotesize sQueeze option 2: commuting the last two gates from the previous decompsition}
    %     \label{fig:myToff2}
    % \end{subfigure}
    \caption{Two equivalent Toffoli decompositions with different basis gates. (a) Qiskit native. (b) sQueeze decomposition, the final two gates commute admitting two different implementations of the Toffoli gate, which may be used to minimise routing.}
\end{figure}
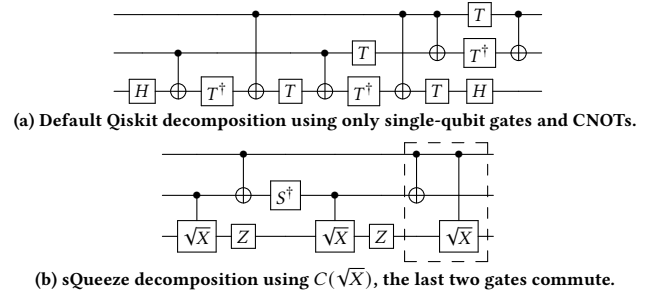

\textit{ }We additionally consider one common three Qubit gate: the Toffoli~\cite{nielsenchuang2011}. As shown in Fig.~\ref{fig:qiskitToffoli}, Qiskit's implementation of the Toffoli gate requires six two-qubit gates and single-qubit rotations~\cite{equivalenceLibrarySourceCode}. By constructing a $C(\sqrt{X})$ gate from the direct $R_{zx}(\theta)$ basis gate the Weyl decomposition gives an implementation of a Toffoli gate that only requires five two-qubit gates, and no additional physical single-qubit gates as shown in Fig.~\ref{fig:myToff1}.

The Toffoli gate poses an additional problem; while the circuit presents operations between all pairs of qubits, this is often not physically realisable on NISQ device topologies~\cite{sabre}. To resolve this dilemma qubits must be swapped to remove non-physical operations, with each swap requiring three CNOTs. The number of these swap operations depends on the allocation of the computational qubits to physical qubits on the device. As a result the total schedule duration for Toffoli gates ranges from $8,544\, \mathtt{dt}$ to $56,192\, \mathtt{dt}$ with a mean of $22,588\, \mathtt{dt}$ depending on the qubits and device topology. As the final two gates of the sQueeze Toffoli implementation commute, it is possible to select which Toffoli will minimise the number of swap operations, further reducing the number of two-qubit gates.

% With the direct Controlled-$\sqrt{X}$ from the Weyl decomposition there is a decomposition of the Toffoli gate which only uses five two-qubit gates instead of six. This is drawn in Fig.~\ref{fig:myToff1}. The last two controlled gates in Fig.~\ref{fig:myToff1} commute, resulting in a third decomposition with the last gates swapped. Theoretically the last two circuits have the same circuit depth but in practice the constraints of hardware mean the schedule lengths may differ. If the three physical qubits were all connected, the depth of circuits in Figures \ref{fig:myToff1} would always be shorter than the decomposition in Fig.~\ref{fig:qiskitToffoli}. Otherwise, a number of SWAP gates must be inserted to make the circuit hardware compliant. The qubits participating in each controlled gate are ordered differently in each decomposition. Consequently, the shortest decomposition pulse schedule depends on the mapping of logical qubits to physical qubits and the device coupling map and we use duration as a heuristic to approximate accuracy.

\subsection{Pass Manager} \label{model:passManager}
Armed with the parameterised pulses in Sec.~\ref{model:1qSection} and \ref{model:2qSection} and their associated gate decompositions in Sec.~\ref{model:equivRelationSection}, sQueeze combines these transformations into a single transpilation function shown in Algorithm~\ref{alg:passManagerAlgo}. First, the circuit is unrolled into the basis gates [$R_x, R_z, R_{zx}, \text{CNOT}$] using Qiskit's existing equivalence library and the new equivalences derived in Sec. \ref{model:equivRelationSection}. Then other passes like routing, layout, and optimisation are applied by using Qiskit's existing \verb:transpile: function. Finally, custom calibration is added to the $R_x(\theta)$ and $R_{zx}(\theta)$ gates based on the work in Sec.~\ref{model:1qSection} and \ref{model:2qSection} respectively.

\begin{algorithm}[h]
\small
\caption{sQueeze Compilation Pipeline}
\label{alg:passManagerAlgo}
\begin{algorithmic}[1]
    \State \emph{pulse-library} $\gets$ Pulse parameters from query server
    \For{\emph{multi-qubit gate} in \emph{circuit}}
            \State Unroll and Decompose gate with Weyl decomposition
    \EndFor
    \State \emph{sel} $\gets$ Qiskit's equivalence library \cite{equivalenceLibrarySourceCode}
    \State \emph{sel}[$U_3$] $\gets$ Equation~ \ref{eq:u3zxzdecomp}.
    \State \emph{basis-gates} $\gets$ [$R_x, R_z, R_{zx}, \text{CNOT}$]
    \For{\emph{gate} \textbf{in} \emph{circuit}}
        \If {\emph{gate} \textbf{not in} \emph{basis-gates}}
            \State Replace \emph{gate} with circuit equivalence from \emph{sel}.
        \EndIf
    \EndFor
    \State Run Qiskit's \verb:transpile: function for other transpilation passes
    \For{\emph{gate} \textbf{in} \emph{circuit}}
        \If {\emph{gate} = $R_x(\theta)$ or \emph{gate} = $R_{zx}(\theta)$}
            \State Replace \emph{gate} with parameterised pulse from \emph{pulse-library}
        \EndIf
    \EndFor
\end{algorithmic}
\end{algorithm}

\subsection{Calibration Server}\label{model:server}

\begin{figure*}
    \centering
    \footnotesize
    \begin{tikzpicture}[node distance=3em]
        % User
        \node[module, align=center] (userCircuit) {User \\ Circuit};
        
        \node[module, right=of userCircuit] (update) {Update};
        \node[module, right= of update, align=center] (db) {Results \\ Database};
        \node[module, right= of db] (query) {Query};
        
        \node[module, above= of query, dashed] (userTranspile) {\textit{Pass Manager}};
        \node[module, right=of userTranspile] (userSchedule) {Schedule};
        
        % Device
        \node[module, below =of userSchedule, align=center] (deviceQ) {Device \\ Queue};
        
        \node[module, right=of deviceQ,  align=center] (deviceEx) {Device \\ Execution};
        
        \foreach \Z in {0.9, 0.6, 0.3, 0}{
        \node[fit={(deviceQ) (deviceEx)}, draw, inner sep=3mm, xshift=\Z em, yshift=\Z em, fill=white] (deviceBox) {};
        }

        \node[module, below =of userSchedule, align=center, yshift=-1.2em] (deviceQ) {Queue};
        
        \node[module, right=of deviceQ,  align=center] (deviceEx) {Execution};
        
        \node [above=0.3em of deviceEx, xshift=-1.2em] (deviceTitle) {\textit{Quantum Devices}};
        
        \node[module, right=of deviceEx, align=center, xshift=2em] (userResults) {User \\ Results};

        \draw[->] (userCircuit) |- (userTranspile);
        \draw[->] (userTranspile)--(userSchedule);
    
        % Query Server Box 
        \node[draw, fit margins={left=0.4em,right=0.4em,bottom=0.4em,top=0.5em},
        fit={(query) (update) (db) }, 
        dashed, label={[anchor=center, fill=white, xshift=-6em]above:\textit{Query Server}}] (qs) {};
        
        %arrow between boxes
        \draw[<->] (userTranspile)--(query);
        \draw[->] (update) -- (db);
        \draw[->] (db) -- (query);

        % Calibration Server
        \node[module, align=center, below =of deviceEx, yshift=-1em] (calResults) {Calibration Results};
        \node[module, left=of calResults, align=center] (db) {Results Database};
        
        \node[module, left= of db, align=center, xshift=1em] (csCircuits) {Calibration Circuits};
        \node[module] (selection) at (update|-csCircuits) {Selection};
        
        \node[fit margins={left=0.3em,right=0.3em,bottom=0.6em,top=0.8em}, fit={(csCircuits) (selection) (db)}, draw, inner xsep=3mm, inner ysep=3mm, dashed, label={[anchor=center, fill=white, xshift=9.2em]above:\textit{Calibration Server}}] (cs) {};

        \draw[->] (userSchedule)--(deviceQ);
        \draw[->] (deviceQ)--(deviceEx);
        \draw[->] (deviceEx)--(userResults);

        \draw[->] (csCircuits)|- ++(0, 0.85) -| (deviceQ);
        \draw[->] (calResults) --(db);
        \draw[->] (db)|- ++(0, -0.7) -| (selection);
        \draw[->] (selection)|- ++ (0, 1) -| (update);
        \draw[->] (deviceEx) -- (calResults);
        \draw[<-] (csCircuits) -- (selection);
    \end{tikzpicture}
    \captionsetup{aboveskip=1em}
    \caption{Custom compiler flow to convert a high-level quantum circuit to an optimised pulse schedule, which is executed on a quantum device. The pass manager requests the latest pulse parameters from the query server to build custom $R_{x}(\theta)$ and $R_{zx}(\theta)$ gates. The calibration server collects real device data, processes measurement results, and updates the query server with the best parameters.}
    \label{fig:serverDiagram}
\end{figure*}
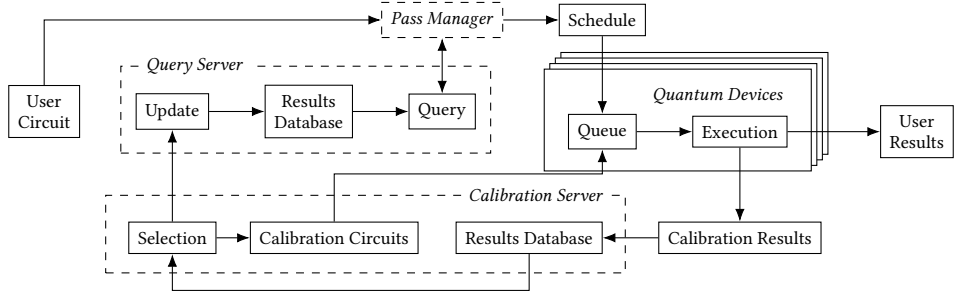
In order to transpile gates down to the pulses described in Sec.~\ref{model:1qSection} and Sec.~\ref{model:2qSection}, we must first characterise these pulses. Rather than block compilation on the completion of these characterisations, we perform these experiments asynchronously on a dedicated `calibration server' and herald the results to a centralised `query server'. The pass manager introduced in Sec.~\ref{model:passManager} may then obtain the most recent pulse parameters from the query server for use in transpiling quantum circuits.  Fig.~\ref{fig:serverDiagram} shows how a high-level quantum circuit is converted to an optimised pulse schedule that can be executed on a NISQ device using our pass manager.

The scheme consists of two parts: a calibration server, and a query server. The calibration server sends experiments to the NISQ backend, processes the measurement results, and selects the best parameters to update the query server. The query server is a fast, lightweight endpoint that stores and returns the selected pulse parameters for $R_x(\theta)$ and $R_{zx}(\theta)$. Each time a circuit is compiled, the sQueeze pass manager requests the current parameters to build optimised pulse schedules.

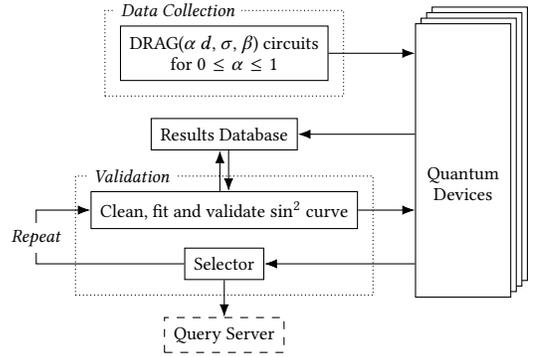
\begin{figure}[h]
    \centering
    \footnotesize
    \begin{tikzpicture}[node distance=1em,
        every node/.style={text centered}, align=center,
        ]
        \tikzset{edge/.style = {->,> = latex'}}
        % Batch
        % \node [module, align=center] (measFitter)     {optimal parameters \\ for qubit $i$} ;
        \node [module, align=center] (ampCircs)                      {DRAG($\alpha$ $d$, $\sigma$, $\beta$) circuits \\ for $0 \leq \alpha \leq 1$};
        \node[fit={(ampCircs)}, draw, inner sep=2mm, densely dotted, label={[anchor=center, fill=white, xshift=-2.5em]above:\textit{Data Collection}}] (batch) {};
        \node [module, below = of ampCircs, yshift=-1em] (db) {Results Database};

        \node [module, below = of db, yshift=-1.2em] (sin2) {Clean, fit and validate $\sin^2$ curve};
        \node [module, below= of sin2]  (selector)   {Selector};
  
        % \node [above= 0.2cm of batch] (label) {\textbf{$R_x$ Calibration}};
        \node[fit={(sin2) (selector) }, draw, inner sep=2mm, densely dotted, label={[anchor=center, fill=white, xshift=-5em]above:\textit{Validation}}] (validationProcess) {};
        
        % Rx flow 
        % \node[fit={(measFitter) (ampCircs) (selector) (validation) (label)  (batch)}, draw, inner xsep=15mm, inner ysep=5mm, dashed] (box) {};
        
        % Device
        \node [right = 4.7em of ampCircs, dashed, align=center, text=white, yshift=0.9em] (device) {DeviceDev};
        
        \node [text=white, below = 12em of device] (emptyCell) {Device};
        % Device box 
        \foreach \Z in {0.9, 0.6, 0.3, 0}{
            \node[fit={(emptyCell) (device)}, draw, inner sep=0mm, xshift=\Z em, yshift=\Z em, fill=white] (box) {};
        }
        \node [below = 6em of device, dashed, align=center] (deviceText) {Quantum \\ Devices};
        
        % Query server
        \node (qs) [module, below = of selector, yshift=-1em, dashed] {Query Server};
        
        % \draw[->]     (measFitter) -- (ampCircs);
        \draw[->]     (selector) -- (qs);
        % node[midway, fill=white, yshift=-2mm] {\emph{Update}};
        % \draw[->]     (db) -- ++(-3, 0) |- (measFitter.west) node [midway, fill=white, , yshift=-1.25cm] {\emph{Repeat}};
        \draw[->]     (selector) -- ++(-2.5, 0) |- (sin2.west) node [midway, fill=white, yshift=-1.5em] {\emph{Repeat}};
        \draw[->]     (ampCircs) -- (box.west |- ampCircs) ;
        \draw[->]     (db.-75) -- (sin2.north -| db.-75);
        \draw[<-]     (db.-105) -- (sin2.north -| db.-105);
        % node [midway, fill=white] {\emph{Query}};

        \draw[->]     (box.west |- db) -- (db.0);
        \draw[->]     (sin2) -- (box.west |- sin2);
        \draw[->]     (box.west |- selector) -- (selector);
    \end{tikzpicture}
    \captionsetup{aboveskip=1em}
    \caption{$R_x(\theta)$ calibration for qubit $q_i$ uses two concurrent processes that share a database. One collects data as outlined in Sec.~\ref{model:1qSection} and the second fits and validates a $\sin^2$ curve.}
    \label{fig:rxCalDiagram}
\end{figure}

The single-qubit pulse calibrations are performed as illustrated in Fig.~\ref{fig:rxCalDiagram}. The calibration server will then clean the data by removing any outliers more than 1.5 standard deviations from the mean and calculate the trailing average using the last two days worth of data. The resulting pulse is validated against Qiskit's $R_x(\theta)$ pulse and is only posted to the query server if it reports a higher mean accuracy. 

To provide real time $R_{zx}\bigl(\frac{\pi}{2} \bigr)$ pulse parameters the calibration server implements particle filtering as presented in Sec.~\ref{model:2qSection}, cycling through the calibrations for each pair of connected qubits. As the IBM job queue is restricted to 100 circuits per submission, each iteration can only test 23 particles along with the previous best round and Qiskit. As Qiskit is included as one of the submissions in the particle filtering method, the resulting pulse parameters are at least as performant as Qiskit. The Qiskit particle is reported as the best particle out of the $25$ approximately $2.3\%$ of the time.

%% file: chapters/methodology.tex
We benchmark sQueeze against the Gokhale~\cite{gokhale2020optimized}, Earnest~\cite{earnest2021pulse} and native Qiskit~\cite{Qiskit} pulse decompositions over IBM's 5 and 7 qubit devices: Belem, Lima, Manila, Quito, and Nairobi. Our primary metrics are the error rate and total execution time reported in units of $\mathtt{dt}$.
% are the accuracy of the implementation, reported as the error rate, and the total execution time, reported in units of $\mathtt{dt}$.

% All tests are run periodically over two weeks with 4000 shots per circuit, resulting in over 2.8 billion total experimental shots. First, we show the improvement for individual gates, and then the gains for the larger Toffoli gate and other common quantum algorithms. All tests use the pass manager for transpilation that retrieves the most accurate pulse parameters from the live server. Our performance is compared to IBM Qiskit and if they differ from Qiskit, Gokhale's and Earnest's methods.

{\bf Individual Gate Benchmarks:} We show the pulses generated by sQueeze correctly implement the intended operations by performing tomography~\cite{greenbaum_tomography,merkel_tomography} in the $X, Y,$ and $Z$ bases and benchmark the performance against other methods. We test our new $R_x(\theta)$, and $R_{zx}(\theta)$ for $\theta$ from 0 to $\pi$, and the CNOT and Toffoli gates built with these basis gates. Each gate is tested in the $2^n$ possible states for each basis and all possible qubit allocations~\cite{howard,Granade_2017}.

{\bf Common Quantum Algorithm Benchmarks:} sQueeze is tested against four well-known quantum algorithm benchmarks; the quantum Fourier transform (QFT)~\cite{weinstein2001implementation}, Quantum Approximate Optimization Algorithm (QAOA)~\cite{farhi2014quantum}, the hidden bitstring variant of the Bernstein–Vazirani algorithm (BV)~\cite{bernstein1997quantum}, and the CDKM Ripple Carry Adder~\cite{cuccaro2004new}. These algorithmic benchmarks represent a range of expected workloads and by association commonly used quantum gates.

{\bf Randomised Benchmarking (RB):}\label{method:rb}
We additionally perform randomised benchmarking (RB) trials to compare the average gate fidelity~\cite{knill2008randomized}. RB constructs circuits of depth $k$ randomised quantum operations before applying the inverse of each gate in reverse order for a total action of $I$ on all qubits. In a system without noise, the set of $n$ qubits should return to the initial state $\ket{0}^{\otimes n}$ with $100\%$ probability. As $k$ varies, the average gate fidelity can be calculated by fitting to
\begin{equation} \label{eq:rbCurveEquation}
    P(\ket{0}^{\otimes n}) = \text{SPAM}_{\alpha}p^k + \text{SPAM}_{\beta}
\end{equation}
where the $\text{SPAM}$ terms absorb the state preparation and measurement errors, $\epsilon_{\text{per gate}} = 1 - p - \frac{1 - p}{2^n}$ is the average error rate per gate, and $n$ is the number of qubits in the circuit. We performed two randomised benchmarking trials: the first with the basis set SU(2) to test the DRAG pulse scaling and then on the expanded set SU(4) to test cross-resonance pulse scaling.

%% file: chapters/evaluation.tex
\begin{table*}
    \centering
    \footnotesize
    \sisetup{separate-uncertainty}
    \begin{tabular*}{\linewidth}{p{8em} @{\extracolsep{\fill}}
    S[detect-weight, mode=text, table-format = 2.2(4), table-column-width = 6em]
    S[detect-weight, mode=text, table-format = 2.2(4), table-column-width = 6em]
    S[detect-weight, mode=text, table-format = 2.2(4), table-column-width = 6em]
    S[detect-weight, mode=text, table-format = 2.2(4), table-column-width = 6em]
    S[detect-weight, mode=text, table-format = 1.2, table-column-width = 6em]
    S[table-format = 1.2, table-column-width = 6em]
    S[detect-weight, mode=text, table-format = 1.2, table-column-width = 6em]}
    \toprule
    \multirow{2}{*}{{Benchmark}} & \multicolumn{4}{c}{{Mean Benchmark Error (\%)}} & \multicolumn{3}{c}{{Improvement vs. Qiskit}} \\ \cmidrule(lr){2-5} \cmidrule(lr){6-8}
     &  {sQueeze} &    {Earnest} &  {Gokhale} &  Qiskit &  {sQueeze} &  {Earnest} &  {Gokhale} \\
    \midrule \midrule
    $R_{x}(\theta)$      & \ubold 0.69 (0.13)      & {\footnotesize{N/A}$^\dag$}           & 1.24 (0.50)          & 1.52 (0.60 )      & \ubold 52.70  \ppercent & {\footnotesize{N/A}$^\dag$}  & 11.73  \ppercent \\
    CNOT                 & \ubold 2.87 (0.95)        & {\footnotesize{N/A}$^\dag$}            & {\footnotesize{N/A}$^\dag$}            & 2.88 (0.93)      & \ubold 0.52\ppercent & {\footnotesize{N/A}$^\dag$}    & {\footnotesize{N/A}$^\dag$}    \\
    $R_{zx}(\theta)$     &  \ubold 2.14 (0.52)       & 2.15 (0.47)      & {\footnotesize{N/A}$^\dag$}            & 2.76 (0.83)       & \ubold 22.63 \ppercent & 22.09  \ppercent & {\footnotesize{N/A}$^\dag$}    \\
    Toffoli              & \ubold 3.92 (0.31)    & {\footnotesize{N/A}$^\dag$}            & {\footnotesize{N/A}$^\dag$}        & 4.65 (0.18)      & \ubold 15.80 \ppercent & {\footnotesize{N/A}$^\dag$}    & {\footnotesize{N/A}$^\dag$} \\ \midrule
    RB SU(2)       & 12.40  (0.25)       & {\footnotesize{N/A}$^\dag$}            & \ubold 12.39 (0.24)       & 12.45  (0.24)   & 0.40 \ppercent & {\footnotesize{N/A}$^\dag$}    &  \ubold 0.48 \ppercent \\
    RB SU(4)  & \ubold 4.42 (0.45)         & 4.64 (0.50)           & 6.14  (0.76)          & 5.83 (0.63)        & \ubold 24.25 \ppercent & 20.41 \ppercent & -5.30 \ppercent \\ \midrule
    QAOA                 & \ubold 40.53 (14.61) & 48.22 (21.29) & 52.08 (22.69) & 52.82 (16.09) & \ubold  23.27 \ppercent  & 8.70 \ppercent & 1.16 \ppercent \\
    BV                   & \ubold 67.41 (2.89) & 67.49 (3.03) & 69.01 (3.66) & 68.94 (3.78) & \ubold 2.22 \ppercent  & 2.10 \ppercent & -0.11 \ppercent \\
    QFT                  & 25.81 (9.52) & 26.09 (9.86) & 26.00 (9.78) & \ubold 25.68 (10.05) & -0.51 \ppercent & -1.60 \ppercent & -1.26 \ppercent \\
    CDKM Adder          & \ubold 29.26 (15.99) & 29.64 (16.18) & 29.31 (16.31) & 29.25 (16.39) & \ubold 0.04 \ppercent  & -1.31 \ppercent & -0.18 \ppercent \\
    \bottomrule
    \end{tabular*}
    \caption{Average fidelity performance across all benchmarks. The error is the distance from the ideal probability distribution, except for RB where it is the fitted error per gate. $\dag$ indicates this method is equivalent to Qiskit.}
    \label{tab:errorOverview}
\end{table*}
 
\begin{table*}
    % Duration summary table 
    \footnotesize
    \centering
    \sisetup{separate-uncertainty}
    \begin{tabular*}{\linewidth}{l @{\extracolsep{\fill}} S[detect-weight, mode=text, table-format = 2.1(2), table-column-width = 11em]
    S[table-format = 2.1(3), table-column-width = 6.5em]
    S[table-format = 2.1(3), table-column-width = 6em]
    S[table-format = 2.1(3), table-column-width = 11em]
    ccc}
    \toprule
    \multirow{2}{*}{Gate}  & \multicolumn{4}{c}{Gate Schedule Duration ($\mathtt{dt}$)} & \multicolumn{3}{c}{Improvement vs. Qiskit} \\ \cmidrule(lr){2-5} \cmidrule(lr){6-8}
     &  {sQueeze} &    {Earnest} &  {Gokhale} &  Qiskit &  {sQueeze} &  {Earnest} &  {Gokhale} \\
    \midrule \midrule
    $X$ / $\sqrt{X}$    & \ubold 77.6 (11.5) & {\footnotesize{N/A}$^\dag$} & {\footnotesize{N/A}$^\dag$} & 160(0.0) & \textbf{2.06 $\times$}  & {\footnotesize{N/A}$^\dag$}  & {\footnotesize{N/A}$^\dag$}  \\ 
    $R_x$ / $U_3$       & \ubold 77.6 (11.5)  & {\footnotesize{N/A}$^\dag$} & 160 (0.0) & 320 (0.0) & \textbf{4.12 $\times$}  & {\footnotesize{N/A}$^\dag$}   & 2.00 $\times$  \\ 
    $CR(\pi/4)$ & \ubold 508(154.7) & {\footnotesize{N/A}$^\dag$} & {\footnotesize{N/A}$^\dag$} & 564(178.2) & \textbf{1.11 $\times$} & {\footnotesize{N/A}$^\dag$} & {\footnotesize{N/A}$^\dag$} \\
    $R_{zx}(\theta)$ & \ubold 1364.6(299.8) & 1472.7(347.5) & {\footnotesize{N/A}$^\dag$} & 3217.8(712.9) & \textbf{2.36 $\times$} & 2.19 $\times$ & {\footnotesize{N/A}$^\dag$} \\
    Toffoli & \ubold 18786.5(2758.5) & {\footnotesize{N/A}$^\dag$} & {\footnotesize{N/A}$^\dag$} & 22277.9(3961.1) & \textbf{1.18 $\times$} & {\footnotesize{N/A}$^\dag$} & {\footnotesize{N/A}$^\dag$} \\
    \bottomrule
    \end{tabular*}
    \caption{Average gate schedule duration in units of $1 \mathtt{dt} = 0.22ns$. $\dag$ indicates this method is equivalent to Qiskit.}
    \label{tab:durationOverview}
\end{table*}

Here we present a range of empirical benchmarks for sQueeze with a focus on individual gate accuracies, common benchmark quantum circuits and random circuits run over two weeks. Tests are run as frequently as queue times permit (in the best case, every few minutes) throughout Qiskit's calibration period without respect to the state of sQueeze's calibration server. The sQueeze calibration runs on average every four hours and at most every two hours. This ensures that tests are not performed immediately following a pulse calibration or only at the end of Qiskit's calibration period. The fidelities for each benchmark are summarised in Table \ref{tab:errorOverview} and the gate durations are compared in Table \ref{tab:durationOverview}. For each benchmark presented within a table we bold the best performing method. The shaded regions in the figures represent a 95\% confidence interval. 

\subsection{Basis Gate Comparisons}\label{res:performance}
To demonstrate the improvements in the performance and accuracy of sQueeze we provide tomographic benchmarks and pulse durations for the $R_x(\theta)$, CNOT, and $R_{zx}(\theta)$ compared to the Earnest, Gokhale, and Qiskit methods. We additionally evaluate the Toffoli decomposition into these basis gates.

\subsubsection{$R_x(\theta)$ and Single-Qubit Gates} \label{results:rxGates}
\begin{figure}[h]
    \centering
    \begin{subfigure}[b]{0.32\linewidth}
        \centering
        \includegraphics[width=\textwidth]{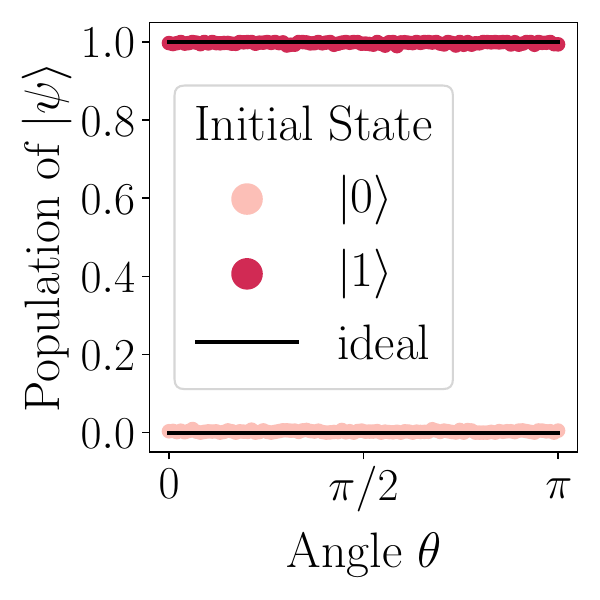}
        \captionsetup{justification=raggedright,margin={2.5em, 0em},singlelinecheck=off}
        \caption{\textit{X} Basis}
        \label{fig:rxbasisX}
    \end{subfigure}
    \begin{subfigure}[b]{0.32\linewidth}  
        \centering 
        \includegraphics[width=\textwidth]{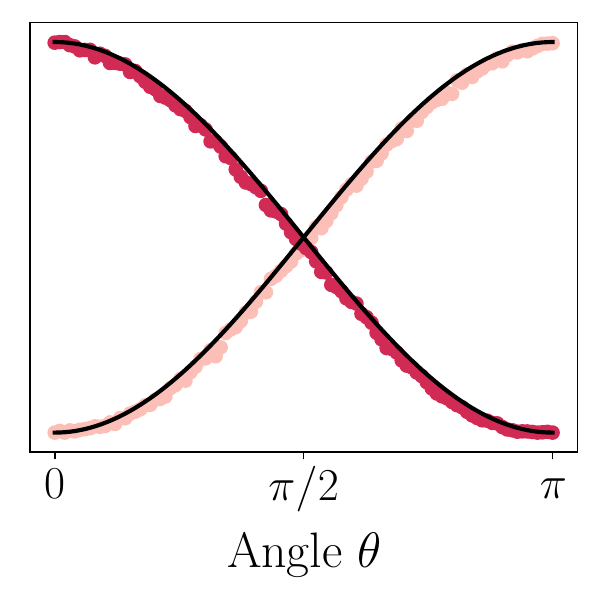}
        \caption{\textit{Y} Basis}
        \label{fig:rxbasisY}
    \end{subfigure}
    \begin{subfigure}[b]{0.32\linewidth}   
        \centering 
        \includegraphics[width=\textwidth]{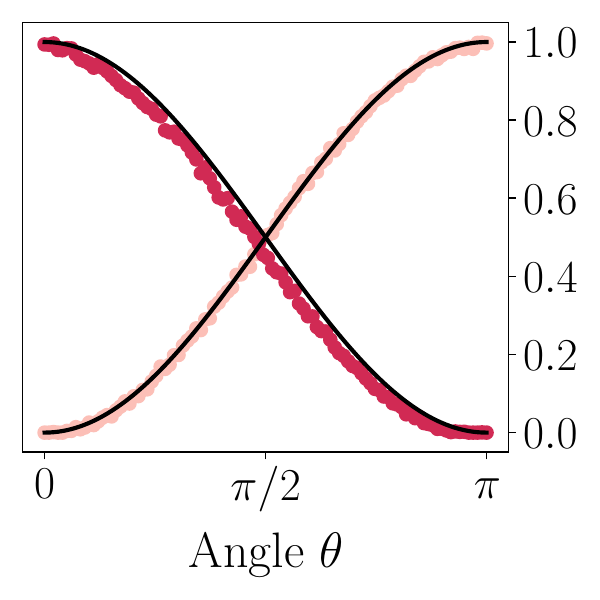}
        \captionsetup{justification=raggedright, margin={1.5em, 0em},singlelinecheck=off}
        \caption{\textit{Z} Basis}
        \label{fig:rxbasisZ}
    \end{subfigure}
    \caption{$R_x(\theta)$ tomography on IBM Quito Qubit 0 over 2.35 million shots and 98 angles in the $X, Y,$ and $Z$ bases.}
    \label{fig:rxTomography}
\end{figure}

\textit{ }To evaluate the accuracy of sQueeze's parameterised $R_x(\theta)$ we perform process tomography for range of $0 \leq \theta \leq \pi$. A representative sample for IBM Quito Qubit 0 over 2.35 million shots is illustrated in in Fig~\ref{fig:rxTomography}. The error for each basis in this trial were 0.3\% (X), 0.8\% (Y), 0.9\% (Z). This is repeated for all qubits on all benchmarks devices and confirms the error is similar in all bases.

\begin{figure}
    \centering
    \includegraphics[width=\linewidth]{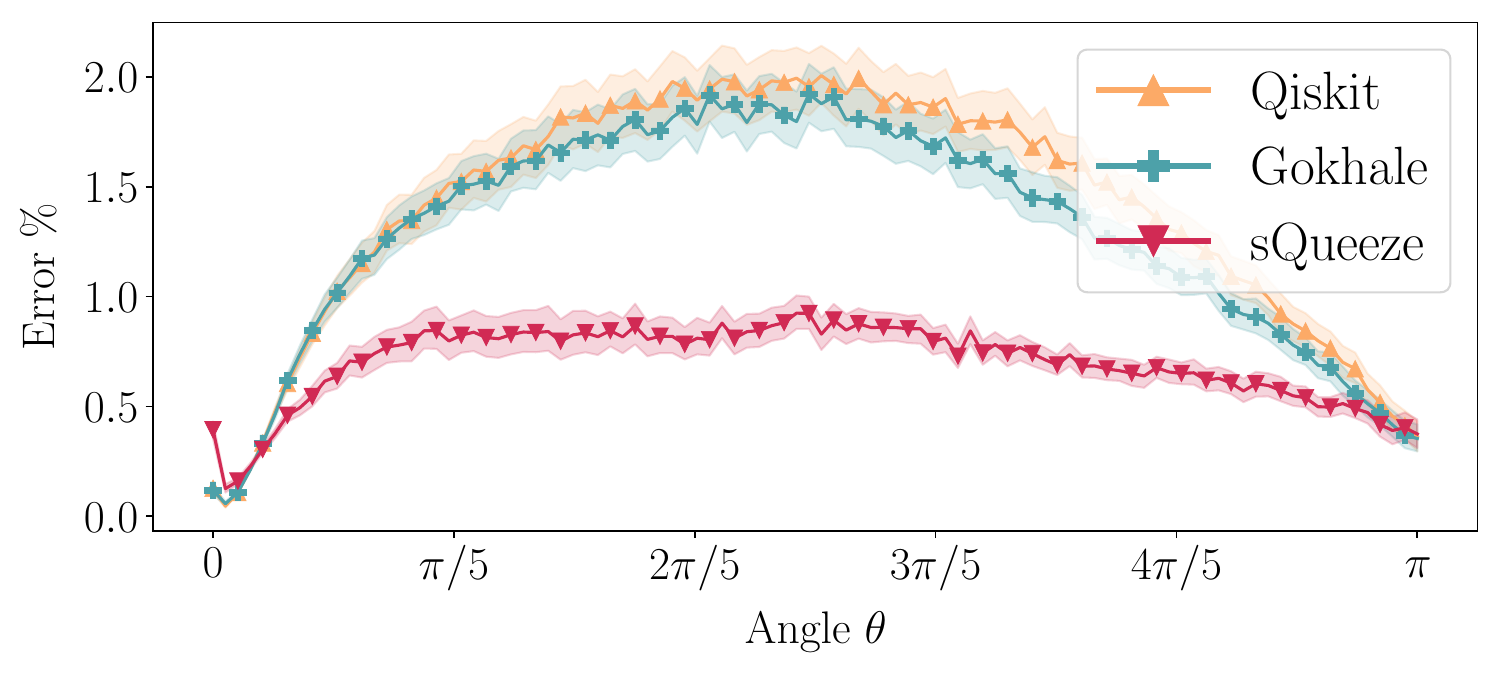}
    
    \caption{$R_x(\theta)$ error for all qubits on the benchmark devices excluding IBM Belem over and 592 million shots.}
    \label{fig:rxResidualerror}
\end{figure}

To compare against other state-of-the-art methods we perform logical $R_x(\theta)$ gates over the range $0 \le \theta \le \pi $ and use each method's respective pulse decomposition. We execute these pulse schedules for the $Z$ basis states and measure the 1-norm distance between the ideal distribution and the measurements results. The average error over all benchmark devices excluding Belem can be seen in Fig.~\ref{fig:rxResidualerror}. We tested each qubit on the benchmark devices continuously for a total of 592 million shots. For single-qubit pulses, we note that the Earnest method is equivalent to Qiskit. sQueeze demonstrates a mean gate error of 0.69\% compared to Gokhale's mean error of 1.24\% and Qiskit's mean error of 1.52\%. sQueeze shows an average $44.3\%$ reduction in the error rate of individual $R_x(\theta)$ gates compared to the next best method. Expanded results for each device can be found in Table \ref{tab:rxDistancePerDevice}, where sQueeze shows up to a 67.1\% improvement compared to the next best method.

\begin{table}[h]
    \centering
    \footnotesize
    \begin{tabular*}{\linewidth}{p{4.5em}@{\extracolsep{\fill}}ccccc}
    \toprule
    \multirow{2}{*}{Device}  & \multicolumn{3}{c}{\footnotesize{Mean $R_x(\theta)$ Error (\%)}} & \multicolumn{2}{c}{\makebox[0pt]{\footnotesize{Improvement vs. Qiskit}}} \\ \cmidrule(lr){2-4} \cmidrule(lr){5-6}
     &  \footnotesize{sQueeze} &   \footnotesize{Gokhale} &  \footnotesize{Qiskit} &  \footnotesize{sQueeze} &  \footnotesize{Gokhale} \\
    \midrule \midrule
    Lima      & \textbf{0.6} \footnotesize{$\pm$ \textbf{0.1}} & 1.8 \footnotesize{$\pm$ 0.5} & 1.8 \footnotesize{$\pm$ 0.9} & \textbf{67.1} \% & -0.3 \%\\
    Manila    & \textbf{0.8} \footnotesize{$\pm$ \textbf{0.7}} & 1.0 \footnotesize{$\pm$ 0.8} & 1.0 \footnotesize{$\pm$ 0.8} & \textbf{24.4} \% & 2.8 \%\\
    Nairobi       & \textbf{0.7} \footnotesize{$\pm$ \textbf{0.3}} & 0.7 \footnotesize{$\pm$ 0.3} & 1.2 \footnotesize{$\pm$ 0.6} & \textbf{46.0} \% & 42.0 \%\\
    Quito     & \textbf{0.9} \footnotesize{$\pm$ \textbf{0.4}} & 1.9 \footnotesize{$\pm$ 0.6} & 2.1 \footnotesize{$\pm$ 1.0} & \textbf{57.9} \% & 8.8 \%\\
    \midrule
    \textit{Average} & \textbf{0.7} \footnotesize{$\pm$ \textbf{0.1}} & 1.3 \footnotesize{$\pm$ 0.6} & 1.5 \footnotesize{$\pm$ 0.5} & \textbf{52.7} \% & 11.7 \% \\
    \bottomrule
    \end{tabular*}
    \caption{The mean $R_x(\theta)$ error per qubit for 98 angles between 0 and $\pi$ over 14 days and 592 million shots.}
    \label{tab:rxDistancePerDevice}
\end{table}

\begin{figure}[h]
    \centering
    \begin{subfigure}[b]{0.49\linewidth}  
        \centering 
        \includegraphics[width=\textwidth]{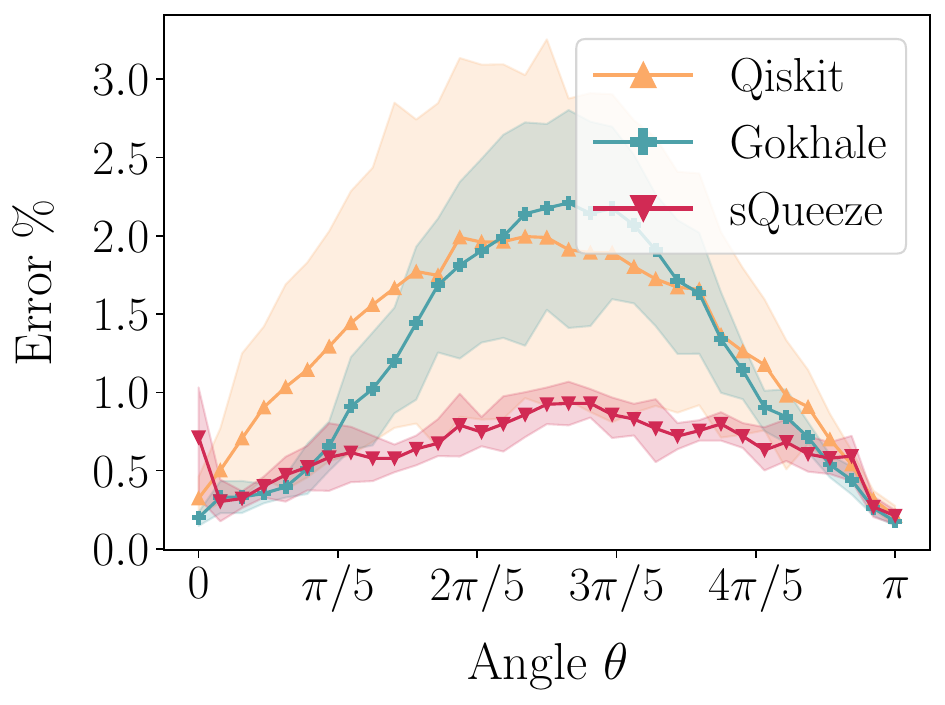}
        \captionsetup{justification=raggedright,margin={3.5em, 0em},singlelinecheck=off}
        \caption{July - August}
        \label{fig:belemAugRxError}
    \end{subfigure}
    \begin{subfigure}[b]{0.49\linewidth}
        \centering
        \includegraphics[width=\textwidth]{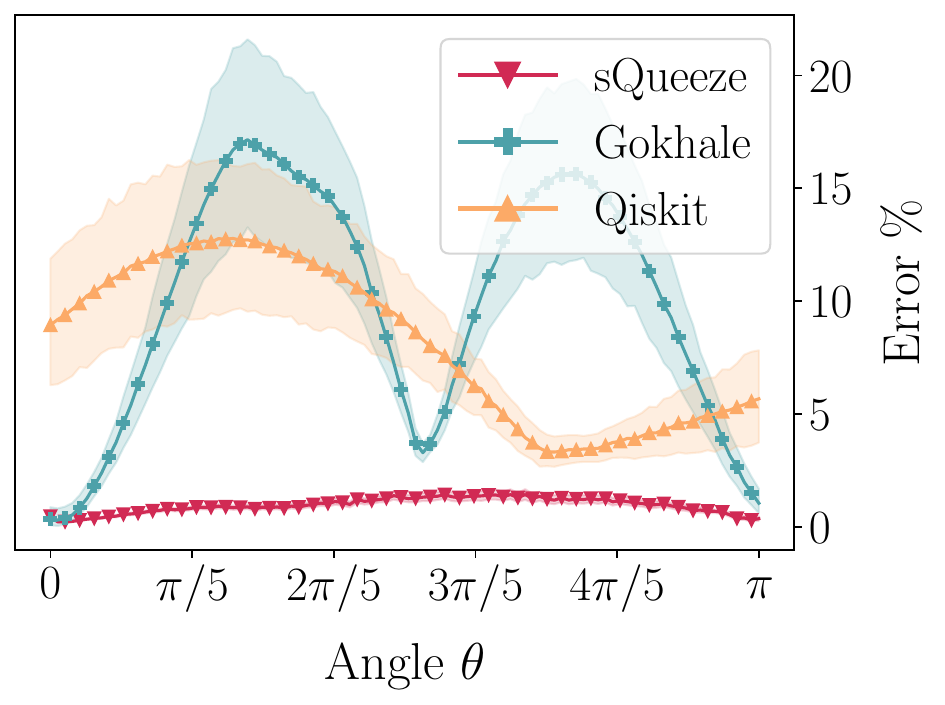}
        \captionsetup{justification=raggedright,margin={1.5em, 1.9em},singlelinecheck=off}
        \caption{September - October}
        \label{fig:belemOctRxError}
    \end{subfigure}
    \caption{A comparison of $R_x(\theta)$ error rates from July to October on the IBM Belem device. Until August the max. average error was 0.9\% (sQueeze), 2.2\% (Gokhale), and 2.0\% (Qiskit) over 109 million shots. By September this jumped to 17.2\% (Gokhale), 12.8\% (Qiskit), while sQueeze remained at 1.42\% over 220 million shots.}
    \label{fig:belemErrorChange}
\end{figure}

Narrowing our focus we examine the performance of a single device that has been excluded from the average in Fig.~\ref{fig:rxResidualerror}. The performance of IBM Belem between July and October 2022 is shown in  Fig.~\ref{fig:belemErrorChange}. This is averaged over all qubits on the device and 309 million experimental shots. Belem was performing similarly to other devices from July to August, but by September the $(3, 4)$ CNOT had a $100\%$ error rate (IBM subsequently removed the connection) and the maximum error $R_x(\theta)$ error jumped from $2.0$ to $12.8\%$. The Gokhale method uses Qiskit's calibration, exacerbating the inaccuracy as the maximum error increases from $2.2\%$ to $17.2\%.$ In comparison, the mean sQueeze error only increased from $0.9\%$ to $1.4\%$, resulting in a mean reduction in error of $88.4\%$ compared to Qiskit and $90.2\%$ relative to Gokhale. This demonstrates that sQueeze can tolerate fluctuation in the calibration of the pulses compared to methods that solely rely on pre-calibrated data. Nonetheless, we have excluded Belem from all subsequent tests for a fair comparison.

Having established the improvement in the accuracy of sQueeze for the $R_x(\theta)$ gate, we next consider the improvement in the corresponding pulse schedule durations. The average schedule durations for 1-qubit gates in Table~\ref{tab:durationOverview} are calculated by using the average $R_x(\theta)$ pulse duration for each qubit listed in Appendix~\ref{appendix:qubitdurations}. sQueeze achieves an average 2.06$\times$ speed-up for the $R_x(\pi)$ and $R_x(\frac{\pi}{2})$ gates that Qiskit directly calibrates.

For more general $R_x$ and $U_3$ gates, sQueeze can implement these gates with a single direct $R_x(\theta)$ instead of two $\sqrt{X}$ gates as in Qiskit's implementation. As a result, sQueeze's mean speed-up doubles to 4.12$\times$ compared to Qiskit. The sQueeze decomposition requires the same number of pulses for a $U_3$ gate as the Gokhale method but with the underlying reduction in the pulse duration sQueeze still exhibits a 2.06$\times$ speed-up. 

\subsubsection{$R_{zx}(\theta)$ and Two-Qubit Benchmarks}\label{res:2q}
\begin{figure}[h]
    \centering
    \includegraphics[width=\linewidth]{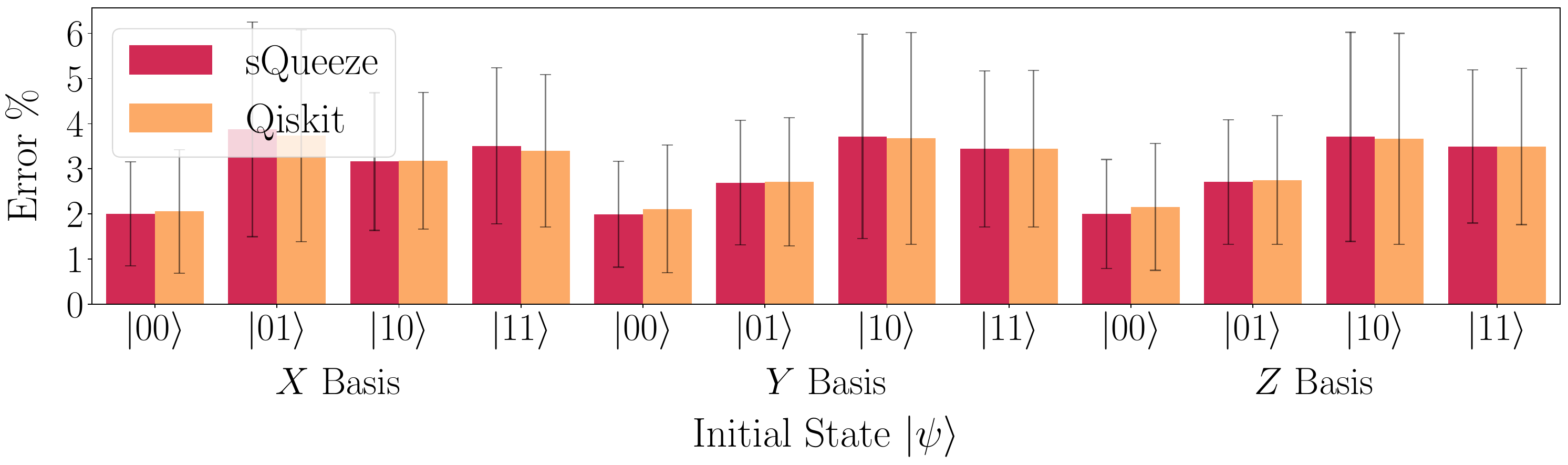}
    \caption{sQueeze CNOT error averaged over all connected qubit pairs on the benchmark devices and 245 million shots.}
    \label{fig:cnotTomog}
\end{figure}

\textit{  }In addition to single-qubit gates, sQueeze also scales the cross-resonance pulses for qubits pairs on IBM devices. These entangling pulses are the basis of all CNOT gates on the benchmark devices. To test the fidelity of the sQueeze CNOT, we use the 1-norm distance between the observed measurement results and the ideal distribution as a metric for the gate error. As the Earnest and Gokhale CNOT implementations are equivalent to Qiskit, we compare the sQueeze CNOT to the Qiskit CNOT over the range of computational basis states shown in Fig.~\ref{fig:cnotTomog}. The measurement results are averaged over 245 million shots and all qubits pairs on the benchmark devices except Belem as discussed above. In Fig.~\ref{fig:cnotTomog}, it is evident that the sQueeze and Qiskit error for each state is approximately equivalent. The average sQueeze CNOT error is 2.87 $\pm$ 0.95\%, which is commensurate with Qiskit's error of 2.88 $\pm$ 0.93\%. This demonstrates that sQueeze achieves an equivalent CNOT fidelity to Qiskit while utilising faster cross-resonance pulses.

To determine the speed-up of the sQueeze CNOT, we take the average of pulse scaling factor $c$ from the particle filter server database and build the corresponding scaled $R_{zx}\bigl(\pm \frac{\pi}{4}\bigr)$ for every pair of connected qubits on the benchmark devices. The sQueeze cross-resonance pulses are up to 1.61$\times$ faster than Qiskit's and 1.11$\times$ faster on average. The calibration server ensures the pulses are never slower than Qiskit. Hence, sQueeze provides faster and equal fidelity CNOTs. 

\begin{figure}[h]
    \centering
    \begin{subfigure}[b]{0.32\linewidth}
        \centering
        \includegraphics[width=\textwidth]{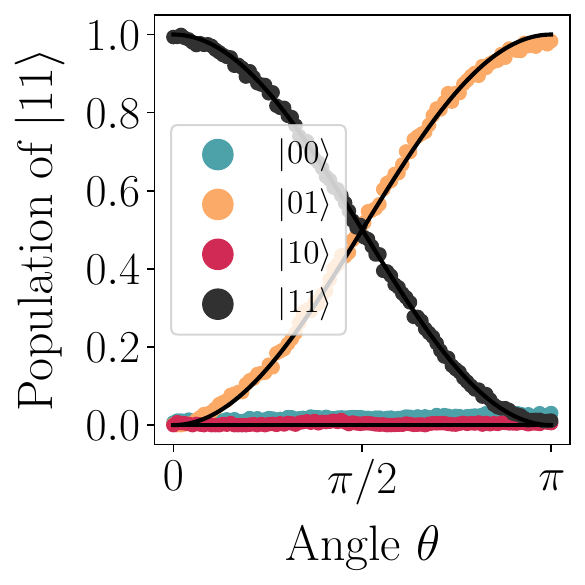}
        \captionsetup{justification=raggedright,margin={3em, 0em},singlelinecheck=off}
        \caption{\textit{X} Basis}
        \label{fig:rzxbasisX}
    \end{subfigure}
    \begin{subfigure}[b]{0.32\linewidth}  
        \centering 
        \includegraphics[width=\textwidth]{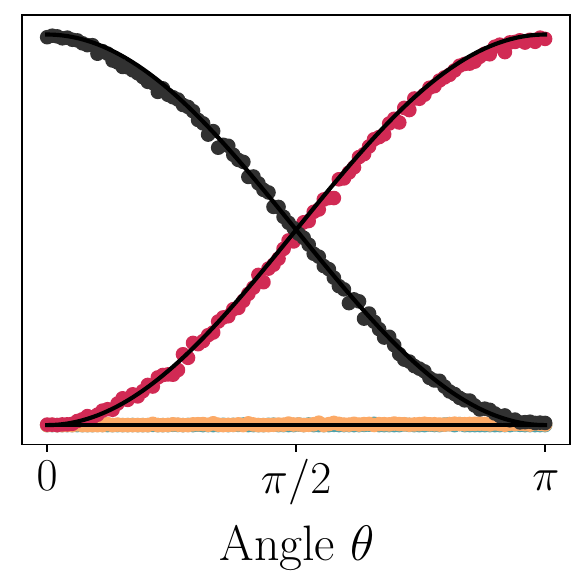}
        \caption{\textit{Y} Basis}
        \label{fig:rzxbasisY}
    \end{subfigure}
    \begin{subfigure}[b]{0.32\linewidth}   
        \centering 
        \includegraphics[width=\textwidth]{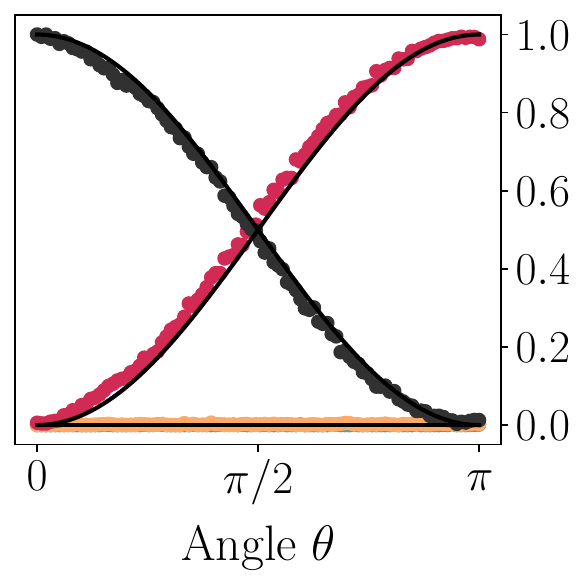}
        \captionsetup{justification=raggedright,margin={1.7em, 0em},singlelinecheck=off}
        \caption{\textit{Z} Basis}
        \label{fig:rzxbasisZ}
    \end{subfigure}
    \caption{$R_{zx}(\theta)$ tomography on IBM Nairobi qubits (0,1) over 1.58 million shots. Mean error 1.00\%($X$), 0.53\%($Y$), 1.03\%($Z$).}
    \label{fig:rzxTomography}
\end{figure}

To create parameterised $R_{zx}(\theta)$ gates, sQueeze uses the Earnest method in combination with the calibration server pulse parameters to generate faster Gaussian-Square pulses. To confirm the fidelity of the new $R_{zx}(\theta)$ gates, we perform tomography over the range $0 \le \theta \le \pi$. This was repeated for all qubits on all benchmark devices to confirm that all bases had similar error. An example for IBM Nairobi Qubits (0,1) is shown in Fig.~\ref{fig:rzxTomography} over 1.58 million shots. The mean error in each basis was $1.00\%$ (X), $0.53\%$ (Y), and $1.03\%$ (Z).

\begin{figure}[h]
    \centering
    \includegraphics[width=\linewidth]{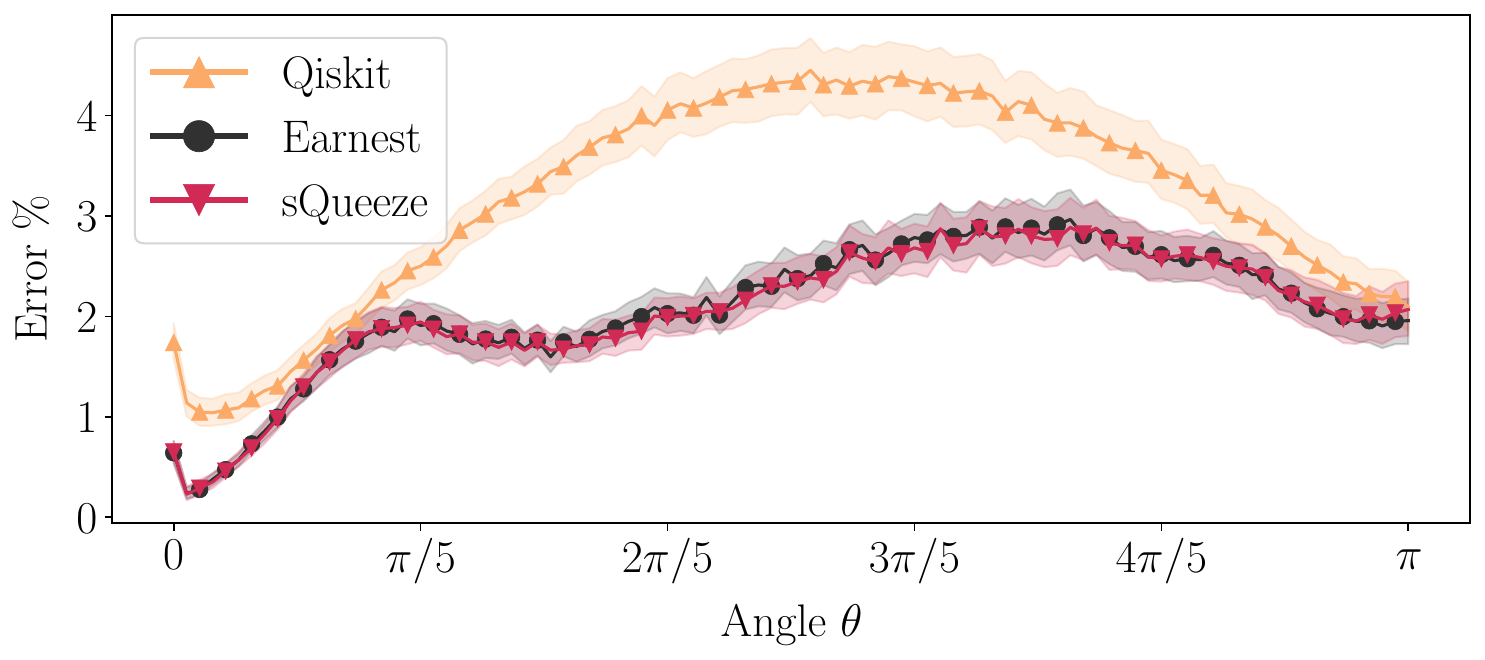}
    \caption{$R_{zx}(\theta)$ error for all connected qubit pairs on all benchmark devices excluding Belem over 14 days and 608 million shots.}
    \label{fig:rzxResidualerror}
\end{figure}

To benchmark the sQueeze $R_{zx}(\theta)$ gates against Earnest and Qiskit we compare average performance in the $Z$ basis states. The 1-norm distances of the measurement results to the action of the ideal operation over the range $0 \le \theta \le \pi$ are shown in Fig.~\ref{fig:rzxResidualerror}. This was averaged over all qubits on all the benchmark devices excluding IBM Belem. The sQueeze pulses have an average error of 2.1\%, compared to 2.2\% for Gokhale and 2.8\% for Qiskit over 608 million shots. This demonstrates sQueeze can reduce error by an average of $22.6\%$ and up to $49.0\%$ for IBM Nairobi relative to Qiskit.  For an expanded set of results per device see Table~\ref{tab:rzxDistancePerDevice}.

\begin{table}[h]
    \centering
    \scriptsize
    \begin{tabular*}{\linewidth}{l @{\extracolsep{\fill}} ccccc}
    \toprule
    \multirow{2}{*}{Device}  & \multicolumn{3}{c}{Mean $R_{zx}(\theta)$ Error (\%)} & \multicolumn{2}{c}{\makebox[0pt]{Improvement vs. Qiskit}} \\ \cmidrule(lr){2-4} \cmidrule(lr){5-6}
     &  sQueeze &   Earnest &  Qiskit &  sQueeze &  Earnest \\
    \midrule \midrule
    Lima      & \textbf{1.8} \scriptsize{$\pm$ \textbf{0.3}} & 1.8 \scriptsize{$\pm$ 0.1} & 2.1 \scriptsize{$\pm$ 1.0} & \textbf{13.7} \% & 12.8 \%\\
    Manila    & \textbf{1.9} \scriptsize{$\pm$ \textbf{1.3}} & 2.0 \scriptsize{$\pm$ 1.4} & 2.2 \scriptsize{$\pm$ 1.3} & \textbf{16.2} \% & 10.9 \%\\
    Nairobi       & 2.0 \scriptsize{$\pm$ 1.1} & \textbf{2.0} \scriptsize{$\pm$ \textbf{1.0}} & 3.9 \scriptsize{$\pm$ 1.8} & 49.0 \% & \textbf{49.1} \%\\
    Quito     & 2.9 \scriptsize{$\pm$ 1.1} & \textbf{2.8} \scriptsize{$\pm$ \textbf{1.1}} & 2.9 \scriptsize{$\pm$ 2.1} & 0.1 \% & \textbf{0.8} \%\\
    \midrule
    \textit{Average} & \textbf{2.1} \scriptsize{$\pm$ \textbf{0.5}} & 2.2 \scriptsize{$\pm$ 0.5} & 2.8 \scriptsize{$\pm$ 0.8} & \textbf{22.6} \% & 22.1 \% \\
    \bottomrule
    \end{tabular*}
    \caption{The average $R_{zx}(\theta)$ averaged over all connected qubits on the device.}
    \label{tab:rzxDistancePerDevice}
\end{table} 

Having established the equivalent fidelity of the sQueeze $R_{zx}$ gate to previous methods, we now demonstrate its reduced pulse duration. The schedule lengths averaged over 100 $R_{zx}(\theta)$ gates for $0 \leq \theta \leq \pi$ and all connected qubits on the benchmark devices are listed in Table \ref{tab:rzxThetaDurations}. The Earnest method of scaling results in a 2.14-2.21$\times$ speed-up relative to Qiskit. As we begin scaling the shorter cross-resonance $R_{zx}\bigl(\pm \frac{\pi}{4}\bigr)$ pulses, sQueeze shows a 2.36$\times$ speed-up over Qiskit and that it is possible to implement accurate gates up to 3.08$\times$ faster. Compared to Earnest, sQueeze achieves an average speed-up of $7\%$ with a maximum observed reduction in the pulse duration of $29.7\%$.

\begin{table}
    \centering
    \scriptsize
    \begin{tabular*}{\linewidth}{l@{\extracolsep{\fill}}cccccc}\toprule
    \multirow{2}{*}{Backend} & \multirow{2}{*}{ Qubit  Pair} &\multicolumn{3}{c}{$R_{zx}(\theta)$ Schedule Duration ($\mathtt{dt}$)} & \multicolumn{2}{c}{\makebox[0pt]{Improvement vs Qiskit}} \\
    \cmidrule(lr){3-5} \cmidrule(lr){6-7}
     &  &  Qiskit & Earnest & sQueeze & Earnest & sQueeze  \\ \midrule \midrule
Lima    & (0,1) & 3072   & 1398.4 &\textbf{1338.9} & 2.20$\times$  & \textbf{2.29}$\times$ \\
        & (2,1) & 3008   & 1368.0  &\textbf{1278.4} & 2.20$\times$  & \textbf{2.35}$\times$ \\
        & (3,1) & 4480   & 2089.3 & \textbf{1867.5}& 2.14$\times$ & \textbf{2.40}$\times$  \\
        & (4,3) & 4672   & 2185.3 &\textbf{2057.3} & 2.14$\times$ & \textbf{2.27}$\times$ \\ \midrule
Manila  & (0,1) & 2816   & 1276.2 & \textbf{1245.4} & 2.21$\times$ & \textbf{2.26}$\times$ \\
        & (1,2) & 4544   & 2121.3 & \textbf{1964.5} & 2.14$\times$ & \textbf{2.31}$\times$ \\
        & (2,3) & 3520   & 1616.6 & \textbf{1491.8} & 2.18$\times$ & \textbf{2.36}$\times$ \\
        & (4,3) & 3008   & 1368.0   & \textbf{1275.2} & 2.20$\times$  & \textbf{2.36}$\times$ \\ \midrule
Nairobi & (0,1) & 2560   & 1154.9 & \textbf{1068.5} & 2.22$\times$ & \textbf{2.40}$\times$  \\
        & (1,3) & 2752   & 1245.4 &\textbf{1183.7} & 2.21$\times$ & \textbf{2.32}$\times$ \\
        & (2,1) & 3840   & 1773.8 &\textbf{1246.1} & 2.16$\times$ & \textbf{3.08}$\times$ \\
        & (5,3) & 2496   & 1125.4 & \textbf{1094.7}& 2.22$\times$ & \textbf{2.28}$\times$ \\
        & (5,4) & 2816   & 1276.2 &\textbf{1183.7} & 2.21$\times$ & \textbf{2.38}$\times$ \\
        & (6,5) & 3072   & 1398.4 & \textbf{1365.4}& 2.20$\times$  & \textbf{2.25}$\times$ \\ \midrule
Quito   & (0,1) & 2432   & 1096.0   & \textbf{1068.5} & 2.22$\times$ & \textbf{2.28}$\times$ \\
        & (1,3) & 3328   & 1523.2 & \textbf{1429.1} & 2.18$\times$ & \textbf{2.33}$\times$ \\
        & (2,1) & 2688   & 1215.4 &\textbf{1155.5} & 2.21$\times$ & \textbf{2.33}$\times$ \\
        & (3,4) & 2816   & 1276.2 & \textbf{1247.7} & 2.21$\times$ & \textbf{2.26}$\times$ \\ \midrule
\textit{Average} &   -    & 3217.8 & 1472.7 & \textbf{1364.6} & 2.19$\times$ & \textbf{2.36}$\times$ \\
    \bottomrule
    \end{tabular*}
    \caption{$R_{zx}(\theta)$ schedule durations based on the average length of 100 $R_{zx}(\theta)$ circuits for $0 \leq \theta \leq \pi$.}
    \label{tab:rzxThetaDurations}
\end{table}

\subsubsection{Toffoli Benchmarks}\textit{ }
Using the new direct $R_{zx}(\theta)$, $R_x(\theta)$, and faster CNOTs, we demonstrate the improvement for the larger Toffoli gate. We test the performance over the $X, Y,$ and $Z$ computational basis states. We test sQueeze over all qubit allocations on the benchmark devices excluding IBM Belem. The errors over the $X$, $Y$, and $Z$ basis states were similar with a mean of $5.1\%$ and range of $3.6-7.1\%$ over 20 million shots.

\begin{figure}[h]
    \centering
    \includegraphics[width=\linewidth]{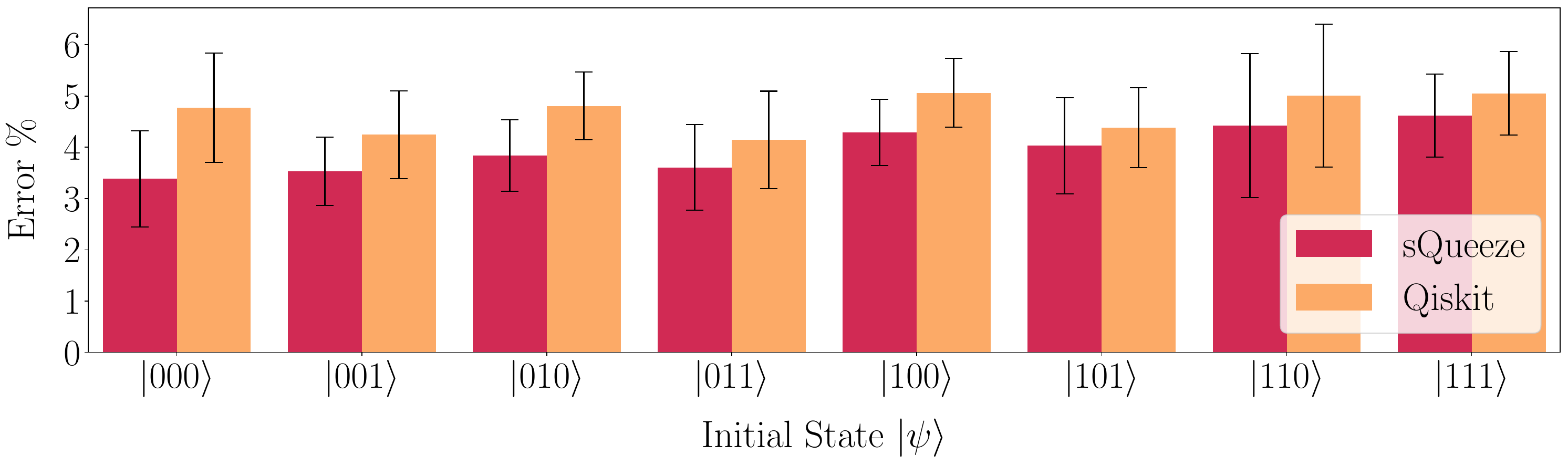}
    \caption{The accuracy of the Toffoli gate, testing all possible qubit allocations for the $Z$ basis over 209 million shots.}
    \label{fig:toffPerformance}
\end{figure}

We additionally compare the performance of sQueeze to current Toffoli implementations by measuring the 1-norm distance from the ideal probability distribution for the full range of $Z$ basis states, noting that Earnest and Gokhale do not provide separate implementations of Toffoli gates. sQueeze demonstrates an average error of $3.92\%$ compared to $4.65\%$ for Qiskit over 209 million shots in Fig.~\ref{fig:toffPerformance}. sQueeze is $15.8\%$ more accurate than Qiskit on average and achieves up to a $29.0\%$ improvement for the $\ket{000}$ state. Moreover, sQueeze achieves better fidelity Toffoli gates than Qiskit for all devices, reaching an average $20.22\%$ improvement for IBM Nairobi as shown in Table \ref{tab:toffoliDistancePerDevice}.

% \begin{figure}
%     \centering
%     \includegraphics[width=\linewidth]{chapters/images/toffoli_tomog.pdf}
%     \caption{Tomography of the Toffoli gate for all possible qubit allocations over 37 million shots.}
%     \label{fig:toffoliTomog}
% \end{figure}

\begin{table}[h]
    \centering
    \scriptsize
    \begin{tabular*}{\linewidth}{lC{7em}C{7em}cc}
    \toprule
    \multirow{2}{*}{Device}  & \multicolumn{2}{c}{Mean Toffoli Error (\%)} & \multirow{2}{5em}{\centering Improvement v. Qiskit} & \multirow{2}{5em}{\centering Speed-up v. Qiskit} \\ 
    \cmidrule(lr){2-3}
     &  sQueeze  &  Qiskit &  \\
    \midrule \midrule
    Lima 	  & \textbf{4.24} \scriptsize{$\pm$ \textbf{0.47}}  & 4.81 \scriptsize{$\pm$ 0.48} & \textbf{11.77\%} & \textbf{1.25}$\times$ \\
    Manila 	  & \textbf{4.04} \scriptsize{$\pm$ \textbf{0.66}} & 4.76 \scriptsize{$\pm$ 0.68} & \textbf{15.22}\% & \textbf{1.16}$\times$\\
    Nairobi 	  & \textbf{3.52} \scriptsize{$\pm$ \textbf{0.11}}  & 4.41 \scriptsize{$\pm$ 0.12} & \textbf{20.22}\% & \textbf{1.13}$\times$\\
    Quito 	  & \textbf{3.87} \scriptsize{$\pm$ \textbf{0.57}} &  4.63 \scriptsize{$\pm$ 0.57} & \textbf{16.39}\% & \textbf{1.18}$\times$\\
    \midrule
    \textit{Average} & \textbf{3.92} \scriptsize{$\pm$ \textbf{0.31}} & 4.65 \scriptsize{$\pm$ 0.18} & \textbf{15.80}\% & \textbf{1.18}$\times$ \\
    \bottomrule
    \end{tabular*}

    % \end{subtable}
    \caption{The mean Toffoli error and duration speed-up based on every possible qubit allocation and 209 million shots.}
    \label{tab:toffoliDistancePerDevice}
\end{table}

We next consider the shorter schedule durations of the higher fidelity Toffoli gates. To compare the schedule durations we generate a Toffoli gate for each possible qubit allocation on all benchmark devices excluding IBM Belem. sQueeze achieves up to a 2.4$\times$ speed-up, namely for Lima qubit allocation (3, 1, 0). The optimal sQueeze schedule is $18,786.5\, \mathtt{dt}$, in contrast to Qiskit's schedule which is $22,277.9\, \mathtt{dt}$ on average. This is a 1.18$\times$ speed-up for sQueeze, as seen in Table~\ref{tab:toffoliDistancePerDevice}. Interestingly, the 15.7\% reduction in schedule duration on average is matched with a 15.8\% decrease in error. Therefore, sQueeze demonstrates that faster and higher fidelity Toffoli gates can be obtained by using schedule duration as a heuristic for accuracy when selecting a Toffoli decomposition.

To use this work without additional routing and mapping methods, the fastest average method per device can be used as the default in place of Qiskit's current implementation. Consequently, Toffoli gates can be sped-up by $1.08\times$ for Lima by simply choosing the decomposition in Fig.~\ref{fig:myToff1} as the default.

\subsection{Quantum Algorithm Benchmarks}\label{res:commonAlgos}
Using sQueeze's higher fidelity and faster gates, we demonstrate that these pulses also work well when used in larger circuits. The fidelity  results are listed in Table~\ref{tab:errorOverview} and for an expanded set of results on a per device basis see Appendix~\ref{tab:summaryAlgosPerDevice}. A particularly strong improvement is exhibited in the QAOA benchmark where sQueeze demonstrates an average $23.27\%$ increase in circuit fidelity over Qiskit. sQueeze demonstrates an approximately equal or better fidelity and the shortest schedule durations for all benchmarks tested.

% \begin{table}
%     \centering
%     \small
%     \begin{tabular*}{\linewidth}{ l@{\extracolsep{\fill}}c}
%     \toprule
%          Benchmark & Number of Gates \\ \midrule\midrule
%          QAOA & $|V|H + |V|P + |E|R_{zz}$ \\ 
%          QFT & $nH + nP + \frac{n(n-1)}{2} CP$\\
%          BV & $2nH + Z + h(s)\text{CNOT}$\\ 
%          CDKM & $(2n + 1)X + (5n + 1)\text{CNOT} + (2n + 1)\text{Toffoli}$ \\ 
%     \bottomrule
%     \end{tabular*}
%     \caption{Number of gates before routing, where $n$ is the size of the bit string input and $V$ and $E$ are the sets of vertices and edges of the input graph for QAOA.}
%     \label{tab:gateEquationsAlgos}
% \end{table}

\subsubsection{Quantum Approximate Optimization Algorithm (QAOA)}\label{res:qaoa}
\begin{figure}
    \centering
    \begin{subfigure}[t]{0.49\linewidth}  
        \centering 
        \captionsetup{width=.9\linewidth}%
        \includegraphics[width=\textwidth]{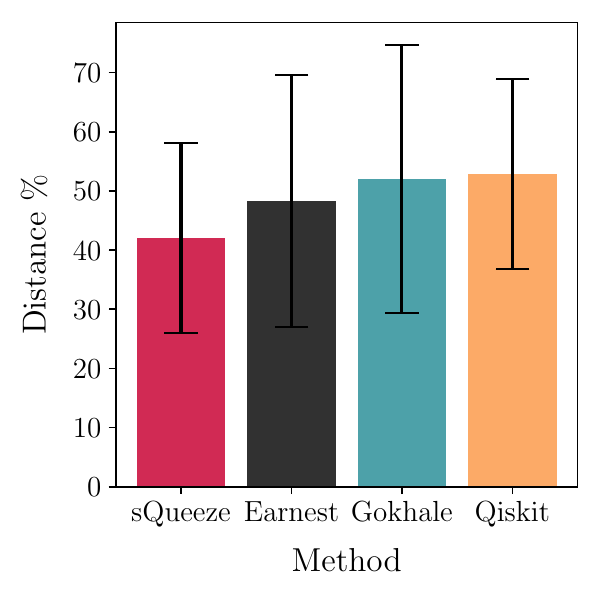}
        \captionsetup{justification=raggedright,margin={4em, 0em},singlelinecheck=off}
        \caption{Mean error}
        \label{fig:qaoaBarChart}
    \end{subfigure}
    \begin{subfigure}[t]{0.49\linewidth}
        \centering
        \captionsetup{width=.9\linewidth}%
        \includegraphics[width=\textwidth]{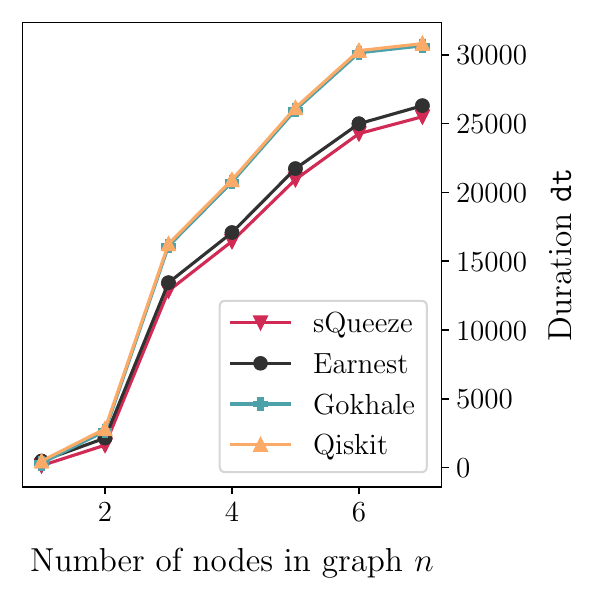}
        \captionsetup{justification=raggedright,margin={0.8em, 0em},singlelinecheck=off}
        \caption{Schedule duration}
        \label{fig:qaoaDuration}
    \end{subfigure}
    \caption{QAOA results over 14.4 million shots and all benchmark devices excluding IBM Belem.}
    \label{fig:qaoaResults}
\end{figure}

We perform circuit fidelity benchmarks between sQueeze and the three other state-of-the-art methods using the 1-norm distance between the measurement results and the ideal probability distribution for each circuit. We test 5 and 7 qubit bipartite input graphs for QAOA to create distinct output probability distributions over all benchmark devices (excepting Belem) for a total 14.4 million shots. For the circuits tested, QAOA requires at least 30 circuit executions per test, and typically ran for 24-30 hours before the classical optimiser terminated. sQueeze demonstrates an average error rate for each circuit execution of $40.53\%$ compared to $48.22\%$ (Earnest), $52.08\%$ (Gokhale) and $52.82\%$ (Qiskit) as seen in Fig.~\ref{fig:qaoaBarChart}. This represents a $22.8\%$ mean improvement compared to Qiskit and up to a $39.6\%$ improvement for the Manila device. This benchmark also demonstrates a $16.4\%$ improvement over Earnest by combining our dynamic $R_x$ and $R_{zx}$ gates, despite recording similar accuracy for the $R_{zx}$ gates as discussed in Sec.~\ref{res:2q}. This demonstrates the benefits of sQueeze's asynchronous calibration and its ability to track drifts in error rates on NISQ devices. 
To quantify the improvement in schedule duration, we compared circuits for connected input graphs with $n-1$ edges for vertices $1 \leq n \leq 7$. The schedule lengths averaged over all benchmark devices are shown in Fig.~\ref{fig:qaoaDuration}.

\subsubsection{Randomised Benchmarking (RB)}
\begin{figure}[h]
    \centering
    \begin{subfigure}[t]{0.49\linewidth}
        \centering
        \includegraphics[width=\textwidth]{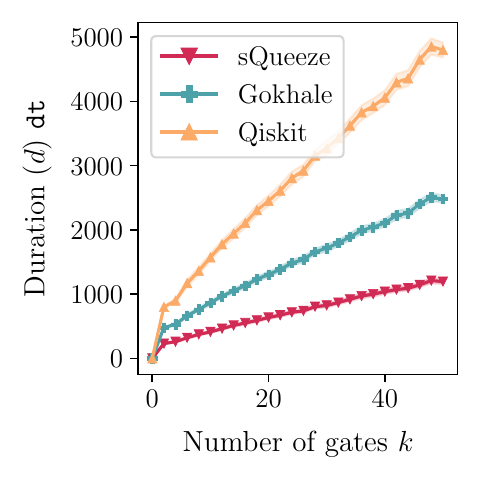}
        \captionsetup{justification=raggedright,margin={4.8em, 0em},singlelinecheck=off}
        \caption{SU(2) gates}
        \label{fig:rb1qDurations}
    \end{subfigure}
    \begin{subfigure}[t]{0.49\linewidth}  
        \centering 
        
        \includegraphics[width=\linewidth]{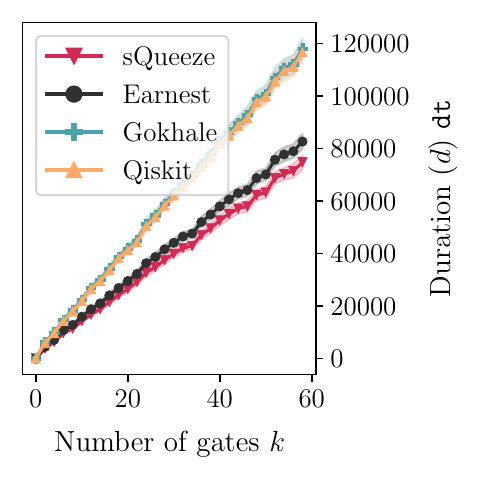}
        \captionsetup{justification=raggedright, margin={1.8em, 0em}, singlelinecheck=off}
        \caption{SU(4) gates}
        \label{fig:rb2qDurations}
    \end{subfigure}
    \caption{RB durations averaged over all benchmark devices.}
    \label{fig:rbDurations}
\end{figure}

\textit{ }To simulate a random circuit and check the performance in an arbitrary basis, we perform randomised benchmarking as described in Sec.~\ref{method:rb}. We test random single-qubit SU(2) gates, and random two-qubit SU(4) gates over all benchmark devices and circuits of size $n$, where $n$ is the number of qubits on the device. For single-qubit SU(2) gates the average gate error rates for each method are $12.40\%$ (sQueeze), $12.39\%$ (Gokhale), $12.45\%$ (Qiskit) showing that all methods perform equally well for SU(2) gates. However, as previously discussed sQueeze SU(2) gates are approximately twice as fast as Gokhale's, which are themselves more than twice as fast as Qiskit. This can be seen in Fig.~\ref{fig:rb1qDurations} where sQueeze achieves a $4.14\times$ speed-up relative to Qiskit and $2.05\times$ speed-up compared to Gokhale.

Performing randomised benchmarking using random SU(4) gates we observe a mean gate error rate of $4.42\%$ (sQueeze), $4.64\%$ (Earnest), $6.14\%$ (Gokhale) and $5.83\%$ (Qiskit) as seen in Table~\ref{tab:summaryAlgosPerDevice}. To measure the average schedule duration, 1000 random RB circuits were generated for every benchmark device and converted into hardware compliant circuits. The schedule durations for these circuits can be seen in Fig.~\ref{fig:rb2qDurations}. For these RB circuits sQueeze achieves a speed-up of $1.56\times$ compared to Qiskit and $1.11\times$ compared to Earnest. 
\subsubsection{Other Quantum Benchmarks}
We test the remaining three benchmarks QFT, BV, and the CDKM ripple carry adder and report similar fidelities. We test all benchmark devices for the circuits specified in Sec.~\ref{section:Methodology} and measure error as the 1-norm distance from the ideal probability distribution, as seen in Table~\ref{tab:errorOverview}. Testing QFT we observe mean errors of 67.4\% (sQueeze), 67.5\% (Earnest), 69.0\% (Gokhale) and 68.9\% (Qiskit). sQueeze demonstrates a 2.2\% reduction in error compared to Qiskit. The BV circuits perform similarly for all methods with average error of 25.8\% (sQueeze), 26.1\% (Earnest), 26.0\% (Gokhale), which are well within one standard deviation of Qiskit's error \todo{$25.7 \pm 10.1$\%}. The same is observed for CDKM, with mean errors of 29.26\% (sQueeze), 29.64\% (Earnest), and 29.31\% (Gokhale), which are again well within one standard deviation of Qiskit's mean error of $29.25 \pm 16.39$\%. An expanded set of results per device can be found in Table~\ref{tab:summaryAlgosPerDevice}.
For these benchmarks, sQueeze produces faster pulse schedules for each circuit, as summarised in Table \ref{tab:durationOverview}. As QFT is dominated by $n/2(n-1)$ controlled-phase gates, sQueeze can achieve a speed-up of 1.34$\times$ over Qiskit for a 7-bit input string. BV is a simple circuit only containing $H$ gates and CNOT gates, for which sQueeze achieves a modest mean speed-up of 1.05$\times$. For the CDKM adders, Qiskit, Earnest, and Gokhale all have equal schedule durations as their $X$, Toffoli and CNOT gates are all equivalent. As the number of Toffoli gates increases by $2n+1$ with each additional adder bit, Toffoli costs dominate the execution time and sQueeze achieves a mean speed-up of 1.10$\times$ for each Toffoli.

%% file: chapters/conclusion.tex
sQueeze demonstrates the benefits of new live calibration techniques which use local searches starting at the device's initial calibrated pulse parameters. By adding parameterised $R_x(\theta)$ and $R_{zx}(\theta)$ we achieve up to a 39.6\% reduction in error for near-term quantum algorithms. We increase the $R_x(\theta)$ fidelity by 52.7\% on average and provide a 4.12$\times$ speed-up compared to standard quantum compilation techniques. $R_{zx}(\theta)$ are 22.63\% more accurate and 2.36$\times$ faster on average. 

The calibration server currently executes experiments for each qubit individually, which is only feasible for small devices. Future work, could investigate sparse calibration techniques that consider the effects of cross-talk when running multiple qubit calibrations in a single circuit. This work supports the importance of new compilation methods that uniquely do not decompose Toffoli gates before qubit mapping and routing~\cite{duckering2021orchestrated}. Future work could extend this to consider the additional Toffoli pulse decompositions we explore here.

%% file: chapters/appendix.tex
\onecolumn
\section{Appendix} 

\subsection{Optimal Single-Qubit Pulse Duration} \label{appendix:qubitdurations}

\begin{table}[H]
    \centering
    \begin{tabular}{lccc} \toprule
    IBM Device & Qubit & Qiskit's $X$ Pulse Amplitude & sQueeze Pulse Duration $\mathtt{dt}$ \\
    \midrule\midrule
    \multirow{5}{*}{Belem}
        & 0 & 0.2511 $\pm$ 0.0049 & 80 \\
        & 1 & 0.2441 $\pm$ 0.0037 & 64 \\ 
        & 2 & 0.0998 $\pm$ 0.0017 & 80 \\ 
        & 3 & 0.2568 $\pm$ 0.0202 & 96 \\ 
        & 4 & 0.3223 $\pm$ 0.0103 & 96 \\ \midrule
    \multirow{5}{*}{Lima}
        & 0 & 0.1199 $\pm$ 0.0023 & 64 \\ 
        & 1 & 0.1465 $\pm$ 0.0022 & 64 \\ 
        & 2 & 0.1373 $\pm$ 0.0024 & 64 \\ 
        & 3 & 0.1262 $\pm$ 0.0022 & 64 \\ 
        & 4 & 0.1261 $\pm$ 0.0020 & 64 \\ \midrule
    \multirow{5}{*}{Manila}
        & 0 & 0.2034 $\pm$ 0.0026 & 80 \\ 
        & 1 & 0.1992 $\pm$ 0.0029 & 80 \\ 
        & 2 & 0.2037 $\pm$ 0.0026 & 80 \\ 
        & 3 & 0.1997 $\pm$ 0.0024 & 80 \\ 
        & 4 & 0.2002 $\pm$ 0.0028 & 80 \\ \midrule
    \multirow{5}{*}{Quito}
        & 0 & 0.1728 $\pm$ 0.0034 & 80 \\ 
        & 1 & 0.1439 $\pm$ 0.0017 & 64 \\ 
        & 2 & 0.1855 $\pm$ 0.0029 & 80 \\ 
        & 3 & 0.1776 $\pm$ 0.0026 & 80 \\ 
        & 4 & 0.4139 $\pm$ 0.0021 & 112 \\ \midrule
    \multirow{7}{*}{Nairobi}
        & 0 & 0.1396 $\pm$ 0.0002 & 64 \\ 
        & 1 & 0.1630 $\pm$ 0.0002 & 80 \\ 
        & 2 & 0.2077 $\pm$ 0.0003 & 80 \\ 
        & 3 & 0.1984 $\pm$ 0.0003 & 80 \\ 
        & 4 & 0.2061 $\pm$ 0.0003 & 80 \\ 
        & 5 & 0.1994 $\pm$ 0.0003 & 80 \\ 
        & 6 & 0.1976 $\pm$ 0.0003 & 80 \\ \midrule
    \textit{Average} & - & - &  77.63 \footnotesize{$\pm$ 11.49} \\ \bottomrule
    \end{tabular}
    \captionsetup{width=.67\textwidth}
    \caption{The optimal sQueeze pulse duration for each qubit. The mean and standard deviation of Qiskit's calibrated $X$ gate amplitude over four months show there is little variation for each qubit.}
\end{table}
\newpage

\subsection{Benchmark Performance per Device} \label{tab:summaryAlgosPerDevice}

\begin{table}[H]
    \centering
    \small
    \begin{tabular*}{\linewidth}{l@{\extracolsep{\fill}} lccccccc}
    \toprule
    \multirow{2}{*}{Benchmark} & \multirow{2}{*}{Device}  & \multicolumn{4}{c}{Mean Benchmark Error (\%)} & \multicolumn{3}{c}{Improvement vs. Qiskit} \\ \cmidrule(lr){3-6} \cmidrule(lr){7-9}
    & &  sQueeze &    Earnest &  Gokhale &  Qiskit &  sQueeze &  Earnest &  Gokhale \\
    \midrule \midrule
    QAOA & Lima 	  & 37.2 \footnotesize{$\pm$ 5.7} & 39.9 \footnotesize{$\pm$ 4.9} & 48.0 \footnotesize{$\pm$ 9.4} & 47.4 \footnotesize{$\pm$ 28.5} & \textbf{21.6} \% & 15.8 \% & -1.2 \%\\
& Manila 	  & 26.0 \footnotesize{$\pm$ 0.0} & 31.8 \footnotesize{$\pm$ 9.4} & 44.0 \footnotesize{$\pm$ 4.5} & 43.1 \footnotesize{$\pm$ 11.6} & \textbf{39.6} \% & 26.2 \% & -2.1 \%\\
& Nairobi 	  & 60.8 \footnotesize{$\pm$ 3.1} & 79.5 \footnotesize{$\pm$ 1.5} & 76.2 \footnotesize{$\pm$ 3.5} & 76.8 \footnotesize{$\pm$ 14.5} & \textbf{20.8} \% & -3.6 \% & 0.8 \%\\
& Quito 	  & 38.0 \footnotesize{$\pm$ 4.3} & 41.6 \footnotesize{$\pm$ 6.4} & 40.2 \footnotesize{$\pm$ 8.8} & 43.9 \footnotesize{$\pm$ 4.8\phantom{0}} & \textbf{13.4} \% & 5.3 \% & 8.4 \%\\ \cmidrule(lr){2-9}
& \textit{Average} & 40.5 \footnotesize{$\pm$ 14.6} & 48.2 \footnotesize{$\pm$ 21.3} & 52.1 \footnotesize{$\pm$ 22.7} & 52.8 \footnotesize{$\pm$ 16.1} & \textbf{23.3} \% & 8.7 \% & 1.2 \% \\ \midrule

RB & Lima 	  & 4.55 \footnotesize{$\pm$ 0.91} & 4.61 \footnotesize{$\pm$ 0.94} & 5.72 \footnotesize{$\pm$ 1.04} & 5.69 \footnotesize{$\pm$ 1.08} & \textbf{20.02} \% & 18.98 \% & -0.51 \%\\
&     Manila 	  & 3.69 \footnotesize{$\pm$ 0.70} & 3.92 \footnotesize{$\pm$ 0.77} & 5.43 \footnotesize{$\pm$ 0.87} & 5.03 \footnotesize{$\pm$ 0.58} & \textbf{26.66} \% & 22.14 \% & -7.92 \%\\
&     Nairobi 	  & 4.73 \footnotesize{$\pm$ 1.04} & 5.30 \footnotesize{$\pm$ 1.09} & 7.76 \footnotesize{$\pm$ 1.84} & 6.90 \footnotesize{$\pm$ 1.34} & \textbf{31.49} \% & 23.11 \% & -12.56 \%\\
&     Quito 	  & 4.84 \footnotesize{$\pm$ 0.42} & 4.88 \footnotesize{$\pm$ 0.52} & 5.84 \footnotesize{$\pm$ 0.52} & 5.89 \footnotesize{$\pm$ 0.50} & \textbf{17.81} \% & 17.16 \% & 0.80 \%\\
    \cmidrule(lr){2-9}
&     \textit{Average} & 4.42 \footnotesize{$\pm$ 0.45}         & 4.64 \footnotesize{$\pm$ 0.50}           & 6.14  \footnotesize{$\pm$ 0.76}          & 5.83 \footnotesize{$\pm$ 0.63}        & \textbf{24.25} \% & 20.41 \% & -5.30 \% \\ \midrule
QFT & Lima 	  & 66.7 \footnotesize{$\pm$ 42.7} & 66.6 \footnotesize{$\pm$ 42.8} & 67.5 \footnotesize{$\pm$ 42.0} & 67.1 \footnotesize{$\pm$ 42.3} & 0.5 \% & \textbf{0.7} \% & -0.6 \%\\
& Manila 	  & 66.2 \footnotesize{$\pm$ 43.4} & 66.4 \footnotesize{$\pm$ 43.2} & 67.8 \footnotesize{$\pm$ 41.8} & 68.0 \footnotesize{$\pm$ 41.8} & \textbf{2.7} \% & 2.4 \% & 0.4 \%\\
& Nairobi 	  & 71.6 \footnotesize{$\pm$ 40.6} & 71.9 \footnotesize{$\pm$ 40.2} & 74.4 \footnotesize{$\pm$ 37.8} & 74.5 \footnotesize{$\pm$ 37.9} & \textbf{3.8} \% & 3.5 \% & 0.1 \%\\
& Quito 	  & 65.1 \footnotesize{$\pm$ 44.5} & 65.1 \footnotesize{$\pm$ 44.5} & 66.4 \footnotesize{$\pm$ 43.1} & 66.2 \footnotesize{$\pm$ 43.4} & 1.6 \% & \textbf{1.6} \% & -0.4 \%\\
\cmidrule(lr){2-9}
& \textit{Average} & 67.4 \footnotesize{$\pm$ 2.9\phantom{0}} & 67.5 \footnotesize{$\pm$ 3.0\phantom{0}} & 69.0 \footnotesize{$\pm$ 3.7\phantom{0}} & 68.9 \footnotesize{$\pm$ 3.8\phantom{0}} & \textbf{2.2} \% & 2.1 \% & -0.1 \% \\ \midrule 

BV & Lima 	  & 18.2 \footnotesize{$\pm$ 13.5} & 18.1 \footnotesize{$\pm$ 13.6} & 18.4 \footnotesize{$\pm$ 13.7} & 18.0 \footnotesize{$\pm$ 13.5} & -1.3 \% & -0.5 \% & -2.0 \%\\
& Manila 	  & 25.2 \footnotesize{$\pm$ 17.5} & 26.1 \footnotesize{$\pm$ 17.4} & 25.3 \footnotesize{$\pm$ 17.5} & 25.2 \footnotesize{$\pm$ 17.1} & \textbf{0.0} \% & -3.7 \% & -0.5 \%\\
& Nairobi 	  & 39.4 \footnotesize{$\pm$ 19.9} & 40.0 \footnotesize{$\pm$ 19.7} & 40.0 \footnotesize{$\pm$ 19.9} & 40.0 \footnotesize{$\pm$ 19.8} & \textbf{1.5} \% & 0.1 \% & 0.1 \%\\
& Quito 	  & 20.4 \footnotesize{$\pm$ 11.6} & 20.2 \footnotesize{$\pm$ 11.3} & 20.3 \footnotesize{$\pm$ 11.5} & 19.5 \footnotesize{$\pm$ 11.2} & -4.6 \% & -3.4 \% & -4.2 \%\\
\cmidrule(lr){2-9}
& \textit{Average} & 25.8 \footnotesize{$\pm$ 9.5\phantom{0}} & 26.1 \footnotesize{$\pm$ 9.9\phantom{0}} & 26.0 \footnotesize{$\pm$ 9.8\phantom{0}} & 25.7 \footnotesize{$\pm$ 10.0} & -0.5 \% & -1.6 \% & -1.3 \% \\ \midrule 
CDKM & Lima 	  & 21.0 \footnotesize{$\pm$ 10.8} & 21.2 \footnotesize{$\pm$ 10.9} & 21.0 \footnotesize{$\pm$ 10.8} & 21.0 \footnotesize{$\pm$ 10.8} & -0.1 \% & -1.0 \% & \textbf{0.1} \%\\
& Manila 	  & 23.8 \footnotesize{$\pm$ 11.9} & 24.3 \footnotesize{$\pm$ 12.0} & 23.7 \footnotesize{$\pm$ 11.7} & 23.3 \footnotesize{$\pm$ 11.6} & -2.2 \% & -4.0 \% & -1.5 \%\\
& Nairobi 	  & 53.1 \footnotesize{$\pm$ 20.6} & 53.7 \footnotesize{$\pm$ 21.8} & 53.6 \footnotesize{$\pm$ 21.8} & 53.7 \footnotesize{$\pm$ 21.8} & \textbf{1.2} \% & -0.0 \% & 0.2 \%\\
& Quito 	  & 19.1 \footnotesize{$\pm$ 9.8\phantom{0}} & 19.4 \footnotesize{$\pm$ 10.0} & 19.0 \footnotesize{$\pm$ 9.8\phantom{0}} & 19.0 \footnotesize{$\pm$ 9.9\phantom{0}} & -0.3 \% & -1.9 \% & \textbf{0.1} \%\\
\cmidrule(lr){2-9}
& \textit{Average} & 29.3 \footnotesize{$\pm$ 16.0} & 29.6 \footnotesize{$\pm$ 16.2} & 29.3 \footnotesize{$\pm$ 16.3} & 29.3 \footnotesize{$\pm$ 16.4} & \textbf{0.0} \% & -1.3 \% & -0.2 \% \\
 \bottomrule
\end{tabular*}
    \caption{Common quantum algorithm benchmark fidelity per device. For RB, we report the fitted mean error per gate, otherwise the error is the distance from the ideal probability distribution. The best method is listed in bold.}
    
\end{table}
\FloatBarrier